\documentclass{aastex}
\usepackage{spr-astr-addons}
\usepackage{url}
\usepackage{layout}
\usepackage{amsmath}
\usepackage{graphicx}
\usepackage{multirow}
\usepackage{xcolor}
\usepackage{times}
\usepackage{natbib}
\usepackage[colorlinks]{hyperref}

\def\aap{A\&A}
\def\aj{AJ}
\def\apss{Ap\&SS}
\def\apj{ApJ}
\def\apjl{ApJL}
\def\apjs{ApJS}
\def\jcap{JCAP}
\def\mnras{MNRAS}
\def\na{New Astronomy}
\def\nar{New Astronomy Reviews}
\def\nat{Nature}
\def\pasp{PASP}
\def\prd{Phys. Rev. D}

\graphicspath{{./figures/}}

\begin{document}

\title{The Cusp/Core problem: supernovae feedback versus the baryonic clumps and dynamical friction model}
\shorttitle{The Cusp/Core problem}
\shortauthors{Del Popolo \& Pace}
\author{A. Del Popolo\altaffilmark{1,2,3}} \and \author{F. Pace\altaffilmark{4}}
\email{adelpopolo@oact.inaf.it}

\altaffiltext{1}{Dipartimento di Fisica e Astronomia, Universit\`a di Catania, Viale Andrea Doria 6, 95125 Catania, 
Italy}
\altaffiltext{2}{INFN sezione di Catania, Via S. Sofia 64, I-95123 Catania, Italy}
\altaffiltext{3}{International Institute of Physics, Universidade Federal do Rio Grande do Norte, 59012-970 Natal, 
Brazil}
\altaffiltext{4}{Jodrell Bank Centre for Astrophysics, School of Physics and Astronomy, The University of Manchester, 
Manchester, M13 9PL, U.K.}

\begin{abstract}
In the present paper, we compare the predictions of two well known mechanisms considered able to solve the cusp/core 
problem (a. supernova feedback; b. baryonic clumps-DM interaction) by comparing their theoretical predictions to 
recent observations of the inner slopes of galaxies with masses ranging from dSphs to normal spirals. We compare the 
$\alpha$-$V_{\rm rot}$ and the $\alpha$-$M_{\ast}$ relationships, predicted by the two models with high resolution 
data coming from \citep{Adams2014,Simon2005}, 
LITTLE THINGS \citep{Oh2015},
THINGS dwarves \citep{Oh2011a,Oh2011b}, 
THINGS spirals \citep{Oh2015},
Sculptor, Fornax and the Milky Way. The comparison of the theoretical predictions with the complete set of 
data shows that the two models perform similarly, while when we restrict the analysis to a smaller subsample of higher 
quality, we show that the method presented in this paper (baryonic clumps-DM interaction) performs better than the one 
based on supernova feedback. 
We also show that, contrarily to the first model prediction, dSphs of small mass could have cored profiles. This means 
that observations of cored inner profiles in dSphs having a stellar mass $<10^6 M_{\odot}$ not necessarily imply 
problems for the $\Lambda$CDM model.
\end{abstract}

\label{firstpage}

\section{Introduction}

The $\Lambda$CDM model is a highly successful paradigm at large scales 
\citep{DelPopolo2007,Komatsu2011,DelPopolo2013,Hinshaw2013,DelPopolo2014,Planck2014_XVI}, but it shows some drawbacks 
at smaller scales (galactic, and centre of galaxy clusters scales 
\citep{DelPopolo2000,DelPopolo2002a,DelPopolo2012c,Newman2013a,Newman2013b,DelPopolo2014})
\footnote{For precision sake, the $\Lambda$CDM paradigm suffers of other problems even at cosmological scales 
(e.g., the cosmological constant problem \citep{Weinberg1989,Astashenok2012}, and the "cosmic coincidence 
problem").}.

Of the main problems of the $\Lambda$CDM paradigm the most "stubborn" seems to be the so called Cusp/Core problem 
\citep{Moore1994,Flores1994}\footnote{The other most often mentioned problems of the $\Lambda$CDM paradigm, are 
a) the discrepancy between the number of subhaloes that N-body simulations predict \citep[e.g.][]{Moore1999} 
and observations; b) the Too-Big-To-Fail (TBTF) problem. In this last problem simulated haloes have too many, too 
dense and massive subhalos in comparison to observations \citep*{BoylanKolchin2011,BoylanKolchin2012}. 
Unified solutions have been proposed to the quoted problems, based on the action of baryons located in the inner 
parts of the haloes \citep{Zolotov2012,DelPopolo2014b}.} dealing with a discrepancy between the flat density 
profiles observed in LSBs and dwarf galaxies, and the cuspy density profile obtained in N-body simulations, e.g. 
the Navarro-Frenk-White (NFW) profile \citep[][]{Navarro1996,Navarro1997,Navarro2010}. The NFW profile predicts an 
inner profile going as $\rho \propto r^\alpha$, with $\alpha=-1$. An even steeper profile predicted by 
\cite{Moore1998} and \cite{Fukushige2001} gives $\rho \propto r^\alpha$, with $\alpha=-1.5$, while other authors 
found that the inner slope is dependent on the object considered, and/or its mass 
\citep{Jing2000,Ricotti2003,Ricotti2004,Ricotti2007,DelPopolo2010,Cardone2011b,DelPopolo2011,DelPopolo2013d,
DiCintio2014}. 
More recent N-body dissipationless simulations seem to agree on the fact that a profile flattening towards the 
centre, to a minimum value of $\simeq -0.8$ \citep{Stadel2009}, namely the Einasto profile, seems to be a better 
fit to simulations \citep{Gao2008}.

The problem is that the smallest value predicted by dissipationless N-body simulations is larger than the values 
obtained by observations 
\citep{Burkert1995,deBlok2003,Swaters2003,KuziodeNaray2011,Oh2011a,Oh2011b}, in SPH simulations 
\citep{Governato2010,Governato2012}, or in semi-analytical models 
\citep{DelPopolo2009,Cardone2012,DelPopolo2012a,DelPopolo2012b,DelPopolo2014a}. {Recently, \cite{Polisensky2015} 
found that haloes do not show universal density profiles, rather their shape is determined by the initial linear 
power spectrum of density perturbations. In addition, the authors found that profiles depend on the halo mass, in 
agreement with previous works on the subject and that warm dark matter (WDM) halos develop a core, but this is not 
significant enough to explain observations.}

The cusp/core problem has been also noticed at galaxy clusters scales. Kinematics and lensing constraints in 
cD galaxies (BCG) located in the centre of relaxed clusters, showed that the clusters DM profiles is flatter than a 
NFW profile, but the total mass profile is in agreement with the NFW predictions 
\citep{Sand2002,Sand2004,Newman2013a,Newman2013b}.

Dwarf galaxies are dark matter (DM) dominated, and have a low baryon fraction \citep{deBlok1997}. They have been 
widely used because of their simple dynamical structure, at least disk galaxies without bulges. In the case of 
high-surface brightness objects (larger objects), it is more complicated to determine the inner density structure. 
So, the previous statement on the cored nature of the inner density profile of all galaxies, is not at all obvious. 
While according to \cite{Spano2008} high-surface brightness galaxies are cored, other authors 
\citep[e.g.,][]{Simon2005,deBlok2008,DelPopolo2012c,DelPopolo2013d,Martinsson2013} conclude differently. The THINGS 
sample shows a tendency to have profiles better described by isothermal (ISO) profiles for low luminosity galaxies, 
$M_B>-19$ and the profiles are equally well described by cuspy or cored profiles for $M_B<-19$. However, even dwarfs 
do not always have flat slopes, as shown by \cite{Simon2005}. In the case of NGC 2976, 4605, 5949, 5693, 6689, the 
authors showed that the profiles range from 0 (NGC2976) to -1.28 (NGC5963). 
Different results have been obtained even using similar techniques for the same object. For example, in the case of 
NGC2976 the dark matter profile slope is bracketed by $-0.17<\alpha<-0.01$, according to \cite{Simon2003}, while 
$\alpha=-0.90 \pm 0.15$ for \cite{Adams2012}, $\alpha=-0.53 \pm 0.14$ according to \cite{Adams2014}, considering 
stars as tracers, or $\alpha=-0.30 \pm 0.18$ \citep{Adams2014}, considering gas as tracer.

The previous discussion highlights the fact that the determination of the inner slope of galaxies, even dwarves, is 
not at all an easy task. The result from the previous studies and several others is that exists a range of profiles, 
and even with the improvements of nowadays kinematic maps there is no agreement on the exact dark matter slopes 
distribution \citep{Simon2005,Oh2011b,Adams2014}.

The situation is even more clear going to larger masses (e.g., spiral galaxies) dominated by 
stars\footnote{See Section~\ref{sect:results} for a wider discussion.}, and especially to smaller masses (e.g. dwarf 
spheroidals (dSphs)) where biases that enter in the system modelling \citep*{Battaglia2013} lead to opposite results.

Several techniques have been used. The spherical Jeans equation gives results highly dependent on the assumptions, 
since mass and anisotropy of the stellar orbits are degenerate in the quoted model \citep{Evans2009}. 
Maximum likelihood in parameter space in Jeans modelling \citep{Wolf2012,Hayashi2012,Richardson2013} has similar 
problems. Schwarzschild modelling has been used for (e.g.) Sculptor and Fornax finding cored profiles 
\citep{Jardel2012,Breddels2013,Jardel2013b,Jardel2013a}. Methods based on multiple stellar populations concluded that 
Fornax (slope measured at $\simeq$ 1 kpc) and Sculptor (slope measured at $\simeq$ 500 pc) have a cored profile 
\citep{Battaglia2008,Walker2011,Agnello2012,Amorisco2012}. 
However, a cusp is found in Draco using a Schwarzschild model \citep{Jardel2013a}. The previous results show that in 
reality there is no accepted conclusion on the inner structures of dSphs. 

On the other side, a clear determination of the cored or cuspy structure of dSphs is very important because the 
smaller is the mass of an object the more probable is that its inner profile is similar to that of dissipationless 
N-body simulations predictions, namely cuspy.

Concerning how the Cusp/Core problem could be solved, there are at least two different approaches: \\
a) cosmological solutions, based on a different spectrum at small scales \citep[e.g.][]{Zentner2003}, 
different nature of the dark matter particles 
\citep{Colin2000,Goodman2000,Hu2000,Kaplinghat2000,Peebles2000,SommerLarsen2001}, or 
modified gravity theories, e.g., $f(R)$ \citep{Buchdahl1970,Starobinsky1980}, $f(T)$ 
\citep[see][]{Bengochea2009,Linder2010,Dent2011,Zheng2011} and MOND \citep{Milgrom1983b,Milgrom1983a}. \\
b) Astrophysical solutions. These are based on the idea that the dark matter component of a galaxy expands due to a 
"heating" mechanism with the result that the inner density is reduced.

The two most known astrophysical solutions are 
a) "supernovae feedback flattening" (SNFF) of the cusp 
\citep{Navarro1996a,Gelato1999,Read2005,Mashchenko2006,Mashchenko2008,Governato2010,Governato2012},\\
and \\
b) "dynamical friction from baryonic clumps" (DFBC) 
\citep{ElZant2001,ElZant2004,Ma2004,Nipoti2004,RomanoDiaz2008,RomanoDiaz2009,DelPopolo2009,Cole2011,Inoue2011,
Nipoti2015}.

A wider discussion of the two models is given in Sections~\ref{sect:sff} and~\ref{sect:gcm}. Here, we may recall 
that the SNFF model in the \cite{Governato2010} results has been found in agreement with the THINGS galaxies 
\citep{deBlok2008,Walter2008} density profiles \citep{Oh2008,Oh2011a,Oh2011b}. Results somehow contradicting the 
previous ones are those of (e.g.) \cite{TrujilloGomez2015} who simulated feedback from supernovae (SN) and radiation 
pressure from massive stars, both in disk galaxies and dwarfs. They found that the second effect is more important 
than supernovae feedback. The ability of the model to solve the small scale problems of the $\Lambda$CDM model has 
also been questioned by several authors 
\citep{Ferrero2012,Penarrubia2012,GarrisonKimmel2013,GarrisonKimmel2014,Papastergis2015} 
(see Section~\ref{sect:sff}).

The DFBC predicted the correct shape of galaxy density profiles \citep{DelPopolo2009,DelPopolo2009a} in agreement 
with the SPH simulations performed by \cite{Governato2010,Governato2012} and of clusters \citep{DelPopolo2012a} by 
\cite{Martizzi2012} and as well predicted correlations among several quantities observed in clusters of galaxies 
\citep{DelPopolo2012a}, later observed in \cite{Newman2013a,Newman2013b}.

Recently, \cite{DiCintio2014} showed that the inner slope depends on the ratio $M_{\ast}/M_{\rm halo}$ in the SNFF 
scheme. A more detailed discussion of the model is performed in Section~\ref{sect:sff}. It predicts a dependence of 
the inner slope from the ratio $M_{\ast}/M_{\rm halo}$. The profile goes from a cuspy one for low values of the 
quoted ratio to cored ones and again to a cuspy one for large spiral galaxies. 
We want to recall that in the DFBC scheme it was found a correlation among the inner slope, the halo mass and the 
angular momentum of the structure \citep{DelPopolo2009,DelPopolo2010,DelPopolo2012a,DelPopolo2012b}, in the case of 
dwarf galaxies, normal spirals, and clusters. In clusters, a series of other correlation were found 
\citep{DelPopolo2012b}, namely a) a correlation among the inner slope and baryonic content to halo mass ratio at 
$z=0$, $M_{\rm b}/M_{500}$, and at $z_{\rm initial}$, $M_{\rm b, in}/M_{500}$ 
\citep[see][their figures 2, 4]{DelPopolo2012b}\footnote{$M_{500}$ is the mass in $R_{500}$, the radius enclosing a 
density 500 times larger than the critical one}, and angular momentum\footnote{Since the angular momentum acquired is 
inversely proportional to the peak height \citep{DelPopolo1996,DelPopolo2009}, the collapse of larger mass structures 
is slowed down}. 
We recall that the baryonic mass $M_{\rm b, in}$ is the initial gas content of the protostructure, while $M_{\rm b}$ 
is the final baryonic content, namely $M_{\rm b}= M_{\rm gas+stars}$. Another confirmation of the goodness of the DFBC 
is that the inner slope dependence on the ratio $M_{\ast}/M_{\rm halo}$ is similar to SPH simulations results, 
although in the case of clusters and for a different mechanism than that of SPH simulations. 
b) In the cluster case, a further correlation between the inner slope and the Brightest Cluster Galaxy 
(BCG) mass, the core radius $r_{\rm core}$ and the effective radius $R_{\rm e}$, and another correlation between the 
mass inside 100 kpc, which is mainly dark matter, and that inside 5 kpc, mainly constituted by baryons 
\citep{DelPopolo2014c} was found \citep[see also][]{DelPopolo2012b}.

Moreover, it was shown that the DFBC mechanism is able to explain the flattening in dwarves as well in clusters 
\citep{DelPopolo2009,DelPopolo2012a,DelPopolo2014}.

In the following, we want to compare the predictions of the SNFF mechanism with the DFBC. To this aim, we will use 
the results of \cite{DiCintio2014}, and compare those results with the observations of \cite{Adams2014}, THINGS 
galaxies, and dwarves, LITTLE THINGS, Sculptor, Fornax, and the Milky Way (MW).

The paper is organized as follows. In Section~\ref{sect:modeldata} we describe the data used and the models that we 
compare to the data. Sections~\ref{sect:results} and~\ref{sect:conclusions} are devoted to results and conclusions, 
respectively. Finally in the appendix \ref{sect:app} we discuss in detail the DFBC mechanism.

\section{Models and data}\label{sect:modeldata}

In the present paper, we will try to discern among the two previously quoted astrophysical mechanisms that are able 
to give rise to cored galaxies. 
Conscious of the limitations in the determination of the inner slope of the density profile $\alpha$, we will use 
the best data nowadays available.

\begin{table*}
 \small
 \centering
 \caption[Galaxy parameters]{Galaxy sample properties from \cite{Adams2014}. The upper (lower) part refers to the 
 gas (star)-tracer data.\label{tab:Adams}}
 \begin{tabular}{@{}lccccccc@{}}
  \tableline
  Galaxy    & $\alpha$           & $\log_{10}M_{\ast}$               & $V_{\rm rot}$              & $\Upsilon_{\ast}$ 
            & $\log_{10}M_{200}$ & $\log_{10}L$ \\
            &                    & $/M_{\odot}$                      & km/s                       &                
            & $/M_{\odot}$       & $/L_{\odot}$ & \\
  \tableline
  NGC 959   & $-0.88\pm 0.15$ & $(0.94_{-0.13}^{+0.12})\times 10^{9}$ & $59.86_{-2.42}^{+2.19}$    & $1.10\pm 0.15$ & 
              $11.06\pm 0.23$ & 8.93 \\
  UGC 2259  & $-0.72\pm 0.09$ & $(0.25_{-0.07}^{+0.06})\times 10^{9}$ & $41.077_{-3.227}^{+2.683}$ & $1.07\pm 0.27$ & 
              $11.42\pm 0.14$ & 8.36 \\
  NGC 2552  & $-0.38\pm 0.11$ & $(1.3_{-0.27}^{+0.2})\times 10^{9}$   & $65.23_{-3.7}^{+3.25}$     & $1.01\pm 0.19$ & 
              $11.33\pm 0.11$ & 9.10 \\
  NGC 2976  & $-0.30\pm 0.18$ & $(0.79\pm 0.21)\times 10^{9}$         & $57.13_{-4.74}^{+3.87}$    & $0.83\pm 0.22$ & 
              $11.94\pm 0.51$ & 8.98 \\
  NGC 5204  & $-0.85\pm 0.06$ & $(0.25\pm 0.03)\times 10^{9}$         & $41.45_{-1.46}^{+1.35}$    & $1.08\pm 0.13$ & 
              $11.36\pm 0.16$ & 8.37 \\
  NGC 5949  & $-0.53\pm 0.14$ & $(2.98_{-0.88}^{+0.92})\times 10^{9}$ & $82.89_{-7.7}^{+6.31}$     & $1.16\pm 0.34$ & 
              $11.82\pm 0.42$ & 9.41 \\
  UGC 11707 & $-0.41\pm 0.11$ & $(1.2_{-0.24}^{+0.3})\times 10^{9}$   & $64.4_{-4.02}^{+3.5}$      & $1.11\pm 0.23$ & 
              $11.49\pm 0.18$ & 9.04 \\
  \tableline
  NGC 959   & $-0.73\pm 0.10$ & $(0.92\pm 0.23)\times 10^{9}$         & $59.55_{-4.62}^{+3.86}$    & $1.08\pm 0.27$ & 
              $11.64\pm 0.32$ & 8.93 \\
  UGC 2259  & $-0.77\pm 0.21$ & $(0.25\pm 0.1)\times 10^{9}$          & $41.4_{-5.54}^{+4.1}$      & $1.10\pm 0.44$ & 
              $11.62\pm 0.61$ & 8.36 \\
  NGC 2552  & $-0.53\pm 0.21$ & $(1.6_{-0.73}^{+0.7})\times 10^{9}$   & $69.11_{-10.5}^{+7.51}$    & $1.24\pm 0.55$ & 
              $11.23\pm 0.38$ & 9.10 \\
  NGC 2976  & $-0.53\pm 0.14$ & $(0.89_{-0.20}^{+0.21})\times 10^{9}$ & $58.98_{-4.09}^{+3.47}$    & $0.93\pm 0.21$ & 
              $11.56\pm 0.46$ & 8.98 \\
  NGC 5204  & $-0.77\pm 0.19$ & $(0.30\pm 0.1)\times 10^{9}$          & $43.67_{-4.54}^{+3.57}$    & $1.30\pm 0.42$ & 
              $11.76\pm 0.51$ & 9.37 \\
  NGC 5949  & $-0.72\pm 0.11$ & $(3.1\pm 0.7)\times 10^{9}$           & $83.68_{-6.02}^{+5.08}$    & $1.20\pm 0.28$ & 
              $11.46\pm 0.22$ & 9.41 \\
  UGC 11707 & $-0.65\pm 0.26$ & $(1.2_{-0.51}^{+0.5})\times 10^{9}$   & $63.78_{-8.82}^{+6.48}$    & $1.07\pm 0.44$ & 
              $11.13\pm 0.37$ & 9.04 \\
  \tableline
 \end{tabular}
\end{table*}

\subsection{Data}\label{sect:data}

The data that we will use are those based on "high resolution integral field spectroscopy" of seven nearby 
galaxies, namely NGC0959, UGC02259, NGC2552, NGC2976, NGC5204, NGC5949, UGC11707, obtained by \cite{Adams2014} 
through measurements of their gas kinematics and integrated stellar light. The complete description of the 
sample selection, photometry, integral field spectroscopy, kinematic extraction of gas and stars are discussed in 
detail in \cite{Adams2014}. The dynamical parameters of the galaxies were obtained using Bayesian statistics, and 
differently from other studies \citep[e.g., dwarf galaxies in][]{Oh2011a,Oh2011b} the entire dark matter density 
profiles were fitted with a Burkert profile \citep{Burkert1995}
\begin{equation}
 \rho(r)=\frac{\rho_{\rm b}}{(1+r/r_{\rm b})(1+(r/r_{\rm b})^2)}\;,
\end{equation}
and a generalized NFW (gNFW) profile \citep{Zhao1996}:
\begin{equation}
 \rho(r)=\frac{\delta_{\rm c}\rho_{\rm c}}{(r/r_{\rm s})^\alpha[1+(r/r_{\rm s})]^{3-\alpha}}\;,
\end{equation}
where
\begin{equation}\label{eqn:deltac}
 \delta_{\rm c}=\frac{200}{3}\frac{c^3}{\zeta(c,\alpha,1)}\;,
\end{equation}
$\rho_{\rm c}$ is the critical density and $c$ the concentration parameter. The function $\zeta(c,\alpha,q_{\rm h})$ 
is defined as \citep{Barnabe2012}:
\begin{equation}
 \zeta(c,\alpha,q_{\rm h})=
 \int_{0}^{c}\frac{\tau^{2-\alpha}(1+\tau)^{\alpha-3}}{\sqrt{1-(1-q_{\rm h}^2)\tau^2/c^2}}~{\rm d}\tau\;,
\end{equation}
where $q_{\rm h}$ indicates the 3D axial ratio of the profile. For spherical symmetry, $q_{\rm h}=1$ as in 
Eq.~\ref{eqn:deltac}.

The complete list of parameters in the study is listed in table~4 of \cite{Adams2014}, and apart the DM parameters, 
the gas-based parameters contain a stellar anisotropy term $\beta_z$ that is used as a free parameter in the "Jeans 
Anisotropic Multi-Gaussian-Expansion" models \citep{Cappellari2008}. 
The parameters were constrained by using gas and stars as tracers.

In table~\ref{tab:Adams}, we report the galaxy name, the slope $\alpha$, $M_{\ast}$, $V_{\rm rot}$, the mass-to-light 
ratio $\Upsilon_{\ast}$, $M_{200}$ and the luminosity $L$. The stellar mass $M_{\ast}$ is calculated from the 
luminosity and $\Upsilon_{\ast}$, while the circular velocity at 2.2 disc scale-lengths $V_{\rm rot}$ is calculated by 
means of the stellar mass Tully-Fisher (TF) relation (Eq.~4 of \cite{Dutton2010}):
\begin{equation}
 \log{\frac{V_{2.2}}{\rm km/s}}=2.143+0.281\left(\log{\frac{M_{\ast}}{10^{10}h^{-2} M_{\odot}}}\right)\;,
 \label{eq:dutt}
\end{equation}
where $V_{2.2}=V_{\rm opt}$. 
We chose this equation because, as we will see later, it was used by \cite{DiCintio2014} in converting $M_{\ast}$ 
to $V_{\rm rot}$ (the rotational velocity) in their figure~6.

Another data set comes from the THINGS dwarfs studied by \cite{Oh2011a,Oh2011b}. The detailed description of the 
parameters of interest to our work and the slope $\alpha$ can be found in \cite{Oh2008,Oh2011a,Oh2011b}. 
Summarizing, they used high-resolution HI data from the THINGS survey
\footnote{THINGS was a HI survey program undertaken with VLA comprising 34 nearby galaxies with high spectral 
($\leq 5.2$ km/s), and spatial ($6^{\prime\prime}$) resolution. 3.6 $\mu$m data, with $4^{\prime\prime}$ resolution, 
from SINGS (Spitzer Infrared Nearby Galaxies Survey) \citep{Kennicutt2003}, were used to constrain the stellar 
component contribution to the total kinematics.}. 
They selected 7 dwarf galaxies with clear rotation pattern from THINGS in order to obtain the rotation curves. In 
order to extract the velocity field from the data cube, different techniques can be used 
(Intensity-Weighted Mean (IWM) velocity field; peak velocity fields; single, multiple Gaussian or Hermite polynomial 
fits). In order to take appropriately into account multiple velocity components, non-circular motions, and so on, 
\cite{Oh2008} introduced a new method to extract from the HI data cube the circularly rotating components, the so 
called "bulk-motion extraction method". 
Since a rotation curve incorporates the dynamics of gas, stars and dark matter, in order to obtain the dynamics of 
the dark matter, it is necessary to extract the baryons contribution from the total dynamics. The stellar component 
mass models are obtained deriving the galaxies luminosity profiles through a tilted ring modelling applied to the 
SINGS 3.6 $\mu$m images to obtain the surface brightness profiles. The luminosity profiles are then converted into 
mass density profiles using a $\Upsilon_{\ast}$ empirical relation, obtained from population synthesis models. 
The HI surface density profile is obtained from the column density of HI, and the tilted-ring model applied to the 
HI maps gives the radial HI distribution. By subtracting the baryons dynamics to the total one, it is possible to 
obtain the dark matter mass model of the galaxies. The halo models used are the NFW and the ISO profiles. 
The last profile is given by
\begin{equation}
 \rho(r)=\frac{\rho_0}{1+(r/r_{\rm c})^2}
\end{equation}
where $r_{\rm c}$ is the core radius, and $\rho_0$ the halo central density.

Using different prescriptions for $\Upsilon_{\ast}$, one can obtain a "maximum disk", a "minimum disk", etc., fit to 
the rotation curve. The density profile can be obtained from the Poisson equation \citep{deBlok2001}:
\begin{equation}
 \rho(R)=\frac{1}{4 \pi G} \left[2\frac{V}{R} \frac{d V}{d R}+\left(\frac{V}{R}\right)^2\right]\;,
\end{equation}
Finally, the inner slope is obtained determining the position where the slope changes most rapidly (break radius). 
A least-squares fit to the inner points to the break radius (usually 5 points) gives $\alpha$. The uncertainty is 
calculated recalculating the slope excluding the data point at the break radius, and including the first point 
outside the break radius. {The error} $\Delta\alpha$ is defined as the difference among these slopes.
Note again that the value of $\alpha$, in this case, was obtained only using the inner points and not the entire 
density profile as in \cite{Adams2014}.

In table~\ref{tab:OhTHINGS}, we show the values of $\alpha$, $M_{\ast}$, and $V_{\rm rot}$ for ICG2574, NGC2366, 
Ho I, Ho II, M81 dwB, DD0153, DD0 154 \citep[see table~1 of][]{Oh2011a}.

\begin{table}
 \small
 \centering
 \caption[Galaxy parameters]{Galaxy sample properties for THINGS dwarfs from \cite{Oh2011a}.\label{tab:OhTHINGS}}
 \begin{tabular}{@{}lcccccc@{}}
  \tableline
  Galaxy      & $\alpha$            & $M_{\ast}$        & $V_{\rm rot}$ \\
              &                     & $(M_{\odot})$         & (km/s)      \\
  \tableline
  IC 2574     & $+0.13\pm 0.07$     & $10.38\times 10^8$    & 77.6 \\
  NGC 2366    & $-0.32\pm 0.10$     & $2.58\times 10^8$     & 57.5 \\
  Holmberg I  & $-0.39\pm 0.06$     & $1.25\times 10^8$     & 38.0 \\
  Holmberg II & $-0.43\pm 0.06$     & $2.00\times 10^8$     & 35.5 \\
  M81 dwB     & $-0.39\pm 0.09$     & $0.30\times 10^8$     & 39.8 \\
  DDO 53      & $-0.38\pm 0.06$     & $0.18\times 10^8$     & 32.4 \\
  DDO 154     & $-0.29\pm 0.15$     & $0.26\times 10^8$     & 53.2 \\
  \tableline
 \end{tabular}
\end{table}

\begin{table}
 \small
 \centering
 \caption[Galaxy parameters]{Galaxy sample properties for LITTLE THINGS \citep{Oh2015}.\label{tab:LITTLE_THINGS}}
 \begin{tabular}{@{}lccc@{}}
  \tableline
  Galaxy    & $\alpha$         & $\log_{10}{M_{\ast}}$         & $V_{\rm rot}$ \\
            &                  & $/M_{\odot}$                  & (km/s)       \\
  \tableline
  DDO 210   & $-0.70\pm 0.04$  & 5.602 & 6.75 \\
  UGC 8508  & $-0.38\pm 0.16$  & 6.477 & 11.9 \\
  CVnIdwA   & $0.03\pm 0.27$   & 6.612 & 12.98 \\
  DDO 216   & $-0.03\pm 1.30$  & 6.934 & 15.98 \\
  WLM       & $0.02\pm 0.02$   & 7.090 & 17.71 \\
  DDO 70    & $-0.48\pm 0.02$  & 7.093 & 17.72 \\
  IC 1613   & $-0.10\pm 0.92$  & 7.288 & 20.14 \\
  DDO 126   & $-0.39\pm 0.05$  & 7.356 & 21.05 \\
  DDO 133   & $-0.11\pm 0.16$  & 7.418 & 21.9 \\
  DDO 168   & $0.62\pm 0.36$   & 7.710 & 26.47 \\
  DDO 101   & $-1.02\pm 0.12$  & 7.730 & 26.81 \\
  HARO36    & $-0.50\pm 0.02$  & 7.764 & 27.33 \\
  DDO 87    & $-0.01\pm 0.48$  & 7.791 & 27.87 \\
  DDO 52    & $-0.49\pm 0.02$  & 7.857 & 29.16 \\
  DDO 50    & $0.10\pm 0.41$   & 7.991 & 31.7 \\
  NGC 2366  & $-0.34\pm 0.10$  & 8.034 & 32.64 \\
  IC 10     & $-0.25\pm 0.32$  & 8.072 & 33.46 \\
  NGC 3738  & $-0.44\pm 0.03$  & 8.096 & 34.89 \\
  NGC 1569  & $-0.23\pm 0.67$  & 8.316 & 39.14 \\
  \tableline
 \end{tabular}
\end{table}

The same technique previously described can be applied to dwarf galaxies from "Local Irregulars Trace Luminosity 
Extremes, The HI Nearby Galaxy Survey" (LITTLE THINGS). In table~\ref{tab:LITTLE_THINGS}, we reproduce $\alpha$, 
$M_{\ast}$, and $V_{\rm rot}$ for a smaller sample of the quoted dwarfs, kindly provided by Se-Heon Oh 
\citep{Oh2015}.

Finally, in table~\ref{tab:spiral_THINGS}, we present $M_{\ast}$ from the THINGS spirals, their $V_{\rm rot}$ 
obtained with the \cite{Dutton2010} formula, and the $\alpha$ values obtained from the rotation curve of the quoted 
galaxies. These data were kindly provided by Se-Heon Oh. Another determination of the slopes of the THINGS galaxies 
was performed by \cite*{Chemin2011} (see Section~\ref{sect:results} for a wider discussion).

\begin{table}
 \small
 \centering
 \caption[Galaxy parameters]{Galaxy sample properties for spiral THINGS (disks) \citep{Oh2015}.
 \label{tab:spiral_THINGS}}
 \begin{tabular}{@{}lccc@{}}
  \tableline
  Galaxy             & $\alpha$  & $\log_{10}M_{\ast}$     & $V_{\rm rot}$ \\
                     &           & $/M_{\odot}$            & (km/s) \\
  \tableline
  NGC7331            & $-1.19$   & $11.26$          & $263.2$\\
  NGC3031            & $-0.80$   & $10.91$          & $209.8$\\
  NGC6946            & $-0.70$   & $10.79$          & $195.4$\\
  NGC3198            & $-0.40$   & $10.49$          & $159.9$\\
  NGC3521            & $-0.10$   & $11.09$          & $235.8$\\
  NGC2403            & $-0.70$   & $9.71$           & $96.5$\\
  NGC7793            & $-0.70$   & $9.44$           & $81.1$\\
  NGC4736            & $-1.0$    & $10.35$          & $146.1$\\
  NGC3621            & $-0.9$    & $10.29$          & $140.5$\\
  NGC2841            & $-1.8$    & $11.13$          & $241.95$\\
  NGC2903            & $-2.0$    & $10.21$          & $133.4$\\
  \tableline
 \end{tabular}
\end{table}

In our analysis we also used the \cite{Simon2005} galaxies not re-studied in \cite{Adams2014} (see 
table~\ref{tab:Simon}). 
\cite{Simon2005} performed an analysis of dwarf and LSB galaxies based on H$\alpha$ high-resolution velocity 
fields for the galaxies NGC 5963, NGC 6689, NGC 4605, and NGC 5949. In the case of NGC 5963 and NGC 4605, CO 
velocity fields were studied. In order to avoid the usual problems connected to the long-slit spectroscopy 
\citep{Simon2003,Swaters2003} in the RCs determination, they used two-dimensional velocity fields. 
Multiple wavelengths velocity fields (e.g., CO, and H$\alpha$) were obtained in order to further reduce systematic 
errors. Multi-color imaging was also used to improve stellar mass-to-light ratio determination. This improves the 
step of modelling and removing the stellar disk. By means of the photometry they measured geometric parameters, and 
used the routine RINGFIT (i.e., tilted-ring modelling) to obtain the radial velocity of the system and RCs starting 
from the velocity fields \citep[see][for a detailed analysis]{Simon2003,Simon2005}.

\begin{table}
 \small
 \centering
 \caption[Galaxy parameters]{Galaxy properties from the \cite{Simon2005} sample.\label{tab:Simon}}
 \begin{tabular}{@{}lccc@{}}
  \tableline
  Galaxy    & $\alpha$        & $M_{\ast}$        & $V_{\rm rot}$\\
            &                 & $(M_{\odot})$     & (km/s)      \\
  \tableline
  NGC4605   & $-0.78\pm 0.04$ & $2\times 10^{9}$  & $74$ \\
  NGC5963   & $-1.20\pm 0.13$ & $9.3\times 10^9$  & $114$ \\
  NGC6689   & $-0.79\pm 0.12$ & $4.5\times 10^9$  & $94$ \\
  \tableline
 \end{tabular}
\end{table}

Finally, we show the density slope $\alpha$, the mass of the stellar component $M_{\ast}$ and the maximum velocity 
$V_{\rm rot}$ for the Milky Way, Fornax and Sculptor in table~\ref{tab:MWFS}.

\begin{table}
 \small
 \centering
 \caption[Galaxy parameters]{Galaxy sample properties for Milky Way \citep{Ascasibar2006}, Fornax and Sculptor 
 \citep{Walker2011}.\label{tab:MWFS}}
 \begin{tabular}{@{}lccc@{}}
  \tableline
  Galaxy    & $\alpha$                & $M_{\ast}$                      & $V_{\rm rot}$\\
            &                         & $(M_{\odot})$                   & (km/s)      \\
  \tableline
  Milky Way & $-1.03\pm 0.04$         & $(6.43\pm 0.63)\times 10^{10}$  & $196.5_{-3.8}^{+3.6}$ \\
  Fornax    & $-0.39_{-0.37}^{+0.43}$ & $(3.12\pm 0.35)\times 10^7$     & $17.8\pm 0.7$ \\
  Sculptor  & $-0.05_{-0.39}^{+0.51}$ & $(8\pm 0.7)\times 10^6$         & $17.3_{-2.0}^{+2.2}$ \\
  \tableline
 \end{tabular}
\end{table}

A further comment regarding the data used is at this point necessary. When we only know the stellar mass 
$M_{\ast}$ of the object considered and we need to infer the rotation velocity $V_{\rm rot}$, we used the relation 
given by \cite{Dutton2010}. We therefore checked the reliability of this relation with objects where both the stellar 
mass and the rotational velocity is known. We noticed though that for some points the velocities inferred differ from 
the true one and therefore few points in the $\alpha-M_{\ast}$ plot are in a different position with respect to the 
same points in the $\alpha-V_{\rm rot}$ plot. Since this is not a systematic effect, we assume nevertheless the 
validity of the \cite{Dutton2010} relation acknowledging the fact that some uncertainties still hold.

\subsection{Models}\label{sect:model}

\subsubsection{Supernovae feedback flattening}\label{sect:sff}

The models we want to compare data with are the two already discussed mechanisms of cusp flattening, namely the 
"supernovae feedback flattening" (SNFF) model 
\citep{Navarro1996a,Gelato1999,Read2005,Mashchenko2006,Mashchenko2008,Governato2010,Governato2012,Teyssier2013}, 
and the "dynamical friction from baryonic clumps" model (DFBC) 
\citep{ElZant2001,ElZant2004,Ma2004,Nipoti2004,RomanoDiaz2008,DelPopolo2009,RomanoDiaz2009,Cole2011,Inoue2011,
Nipoti2015}.

The importance of baryons in solving the Cusp/Core problem was suggested starting from \cite{Flores1994} and 
stressed in many following works. The first mechanism envisaged was connected to supernovae feedback.

\cite{Navarro1996a} showed that the sudden expulsion of baryons into the halo in a single event could flatten 
the profile. However, \cite{Gnedin2002} showed that a single explosive event has not sufficient energy to form a 
core, while repeated moderate violent explosions could reach the goal (however see \cite{GarrisonKimmel2013} 
for a different point of view). 
\cite{Mashchenko2006,Mashchenko2008} showed that in primordial galaxies, random bulk motions of gas driven by SN 
explosions could form a core, and a similar model by \cite{Governato2010} found the same result. 
\cite{Oh2011a,Oh2011b} compared the average slope of THINGS dwarves with the simulations by \cite{Governato2010}, 
and \cite{Governato2012} made a similar comparison for larger objects, and found a correlation among $M_{\ast}$ 
and the inner slope for galaxies having $M_{\ast}>10^6 M_{\odot}$
\footnote{In Galaxies with $M_{\ast}<10^6 M_{\odot}$ the supernovae feedback mechanism was not able to 
transform cusps into cores.}. 
The Governato's papers used the code GASOLINE \citep*{Wadsley2004}, a N-Body+SPH code to simulate galaxies. 
By means of the "zoom" technique \citep{Katz1993}, the resolution for gas particles was 
$M_{\rm p,gas} = 3 \times 10^3 M_{\odot}$, and $M_{\rm p,DM} = 1.6 \times 10^4 M_{\odot}$ for DM particles, and 
the softening 86 kpc. The authors performed a run in which stars formed if the hydrogen density was $> 100/$cm$^3$
(High Threshold (HT) run), and another in which stars formed if the hydrogen density was $> 0.1/$cm$^3$ 
(Low Threshold (LT) run). \\
These simulations, similarly to \cite{DiCintio2014}, implement SN feedback through the blast wave SN feedback 
\citep{Stinson2006}, and/or early stellar feedback \citep{Stinson2013}. Stars with masses larger than $8~M_{\odot}$ 
deposit an energy of $10^{51}$ erg in the interstellar medium (ISM). Even metals are allowed to diffuse 
between the particles of gas \citep*{Shen2010}. The coupling of energy eject from SN to the ISM is obtained 
using a coupling coefficient $\epsilon_{\rm esf}$. In the MaGICC simulations \citep{Stinson2013} the fiducial 
$\epsilon_{\rm esf}=0.1$. In the following (Sect.~\ref{sect:differences}), we will discuss more in detail the 
coupling parameters \citep[e.g.,][]{Penarrubia2012}.

\cite{DiCintio2014}, using the same code (GASOLINE), and similar parameters showed that if 
$M_{\ast}/M_{\rm halo} \leq 0.01\%$ the stellar feedback energy is not enough to turn cusps into cores, and one 
expects cuspy profiles similar to the NFW or more cuspy. Going up with stellar mass (better $M_{\ast}/M_{\rm halo}$) 
the profiles become less steep, and when the ratio $M_{\ast}/M_{\rm halo} \simeq 0.5\%$ one gets the flattest 
profiles. 
For a larger $M_{\ast}/M_{\rm halo}$ ratio the deepening of the potential well of the galaxies, produced by a larger 
number of stars, opposes the SNFF mechanism and galaxies have more cuspy profiles. 
\cite{DiCintio2014} used the stellar mass Tully-Fisher (TF) relation \citep[Eq. 4 of][]{Dutton2010} to have 
predictions on the DM inner slope on the galaxies observed rotation velocity. The model predicts that the galaxies 
in which the cored profiles should be more evident are LSB galaxies 
\citep[in agreement with observations,][]{deBlok2008,Oh2011a}, while in small velocity (mass) dSphs the profiles 
tend to be cuspy. The situation is more complicated for disk galaxies of larger mass (e.g., Milky Way) which are 
baryon dominated and the uncertainties in the disc-halo decomposition are larger. For those galaxies it is difficult 
to have a clear cut on the cuspy or cored nature of the density profile. As already discussed, for larger mass 
galaxies ($V_{\rm rot}>150$ km/s), \cite{deBlok2008} and \cite{Martinsson2013} showed that ISO or NFW fit equally 
well the density profiles. Other studies \citep[e.g.,][]{Donato2004,McGaugh2007} reached the conclusion that cored 
profiles describe well the density profile.

\begin{figure}
 \centering
 \includegraphics[width=\hsize]{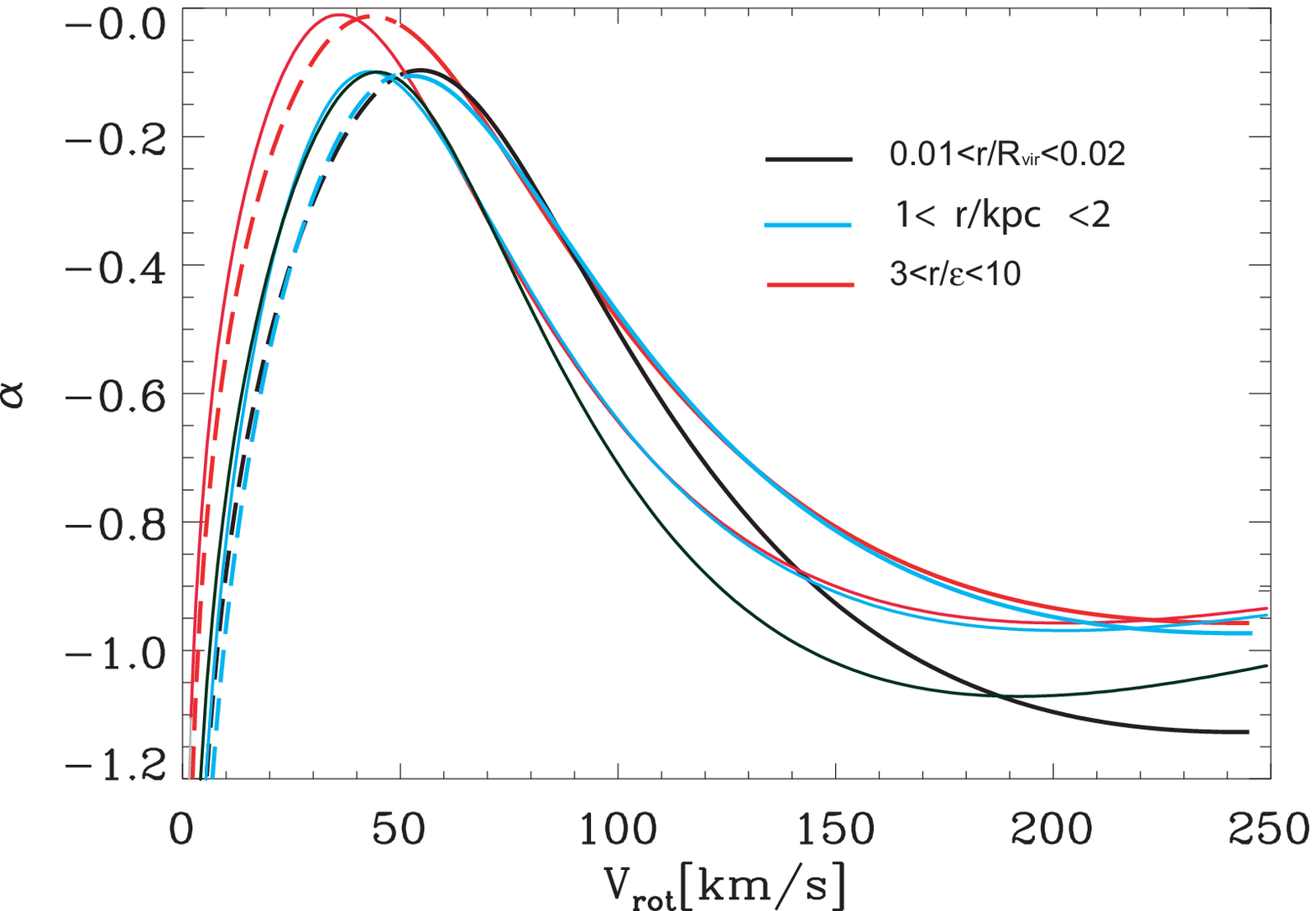}
 \caption[]{The $\alpha$-$V_{\rm rot}$ relation. The dashed lines refer to the calculations of 
 \protect\cite{DiCintio2014}, and the solid lines are those calculated in this paper. The dashed lines and the solid 
 ones are different by a factor h inverse, coming from the Tully-Fisher relation of \protect\cite{Dutton2010}, and 
 discarded in \protect\cite{DiCintio2014}. The black lines refer to the slope calculated at 
 $0.01<r/R_{\rm vir}<0.02$, the cyan lines that calculated at $1<r/{\rm kpc}<2$ and the red lines refer to the slope 
 calculated at $3<r/\epsilon<10$.}
 \label{fig:alphavrot}
\end{figure}

\subsubsection{Gas clumps merging}\label{sect:gcm}

As already reported, the other mechanism able to transform cusps into cores is that proposed by 
\cite{ElZant2001,ElZant2004}\footnote{For precision's sake, the concept that large clouds could heat stellar systems 
was proposed by \cite{Spitzer1951}.}, based on merging gas clumps of $10^{5} M_{\odot}$ in the case of dwarves, and 
$10^{8} M_{\odot}$ in the case of spirals\footnote{In the \cite{Nipoti2015} simulations, the clump mass was 
$10^5-10^6 M_{\odot}$, and the total mass $10^9 M_{\odot}$.}.
Energy and angular momentum transfer from clumps to DM can flatten the profile, and the process is the more 
efficient, the earlier it happened, when halos were smaller. The effectiveness of the process has been confirmed by 
several authors 
\citep{Ma2004,Nipoti2004,RomanoDiaz2008,DelPopolo2009,RomanoDiaz2009,Cole2011,Inoue2011,DelPopolo2014b,Nipoti2015}.

More in detail, as shown in \cite{DelPopolo2009,DelPopolo2014b}, initially the proto-structure is in the linear 
phase, containing DM and diffuse gas. 
The proto-structure expands to a maximum radius and then re-collapses, first in the DM component that forms the 
potential well in which baryons will fall. 
Baryons subject to radiative processes form clumps, which collapse to the centre of the halo while forming stars 
\citep[][(see Sect.~2.2.2,~2.2.3)]{DeLucia2008,Li2010}. In the collapse phase baryons are compressed (adiabatic 
contraction \citep{Blumenthal1986,Gnedin2004}), so making more cuspy the DM profile. 
The clumps collapse to the galactic centre, because of dynamical friction (DF) between baryons and DM, transferring 
energy and angular momentum to the DM component. The cusp is heated, and a core forms, before stars form and stellar 
feedback starts to act expelling a large part of the gas, leaving a lower stellar density with respect to the 
beginning. Feedback destroys clumps\footnote{This is allowed since star formation is a not efficient process.} 
soon after a small part of their mass is transformed into stars. The mass distribution is then dominated by DM.

The model described is in agreement with 
\cite{ElZant2001,ElZant2004,Ma2004,Nipoti2004,RomanoDiaz2008,RomanoDiaz2009,Cole2011,Inoue2011,Nipoti2015}.\\
It is the only model able to explain the correct dependence of the inner slope of the DM profile over 6 order of 
magnitudes in the halo mass, namely from dwarves to clusters 
\citep{DelPopolo2009,DelPopolo2010,DelPopolo2012a,DelPopolo2012b,DelPopolo2014}. 
Here we want to stress another interesting point. 
As previously reported the model showed that the inner slope is mass dependent 
\citep{DelPopolo2009,DelPopolo2010,DelPopolo2012a,DelPopolo2012b}. Moreover in \cite{DelPopolo2012b} and 
\cite{DelPopolo2014c} were found correlations among the inner slope and other quantities with the BCG mass and radius, 
in agreement with results by \cite{Newman2013a,Newman2013b}.

\subsubsection{Drawbacks and differences between the models}\label{sect:differences}
As already described, the SNFF model is based on the expulsion of gas due to SN explosions, while the DFBC model is 
based on dynamical friction between baryons and DM. The two models work in a different way. In the case of the SNFF 
model we start from gas which forms stars which can then explode into SN if they posses the correct mass. So, in this 
model in order to produce the density profile flattening, we need a longer and more complex series of events to 
reach the goal (cusps transformed into cores). In the case of the DFBC, it is enough that big gas clumps are present 
to flatten the density profile. In a few words, the DFBC model is more ergonomic than the SNFF model.

\begin{figure*}
 \centering
 \includegraphics[width=\hsize]{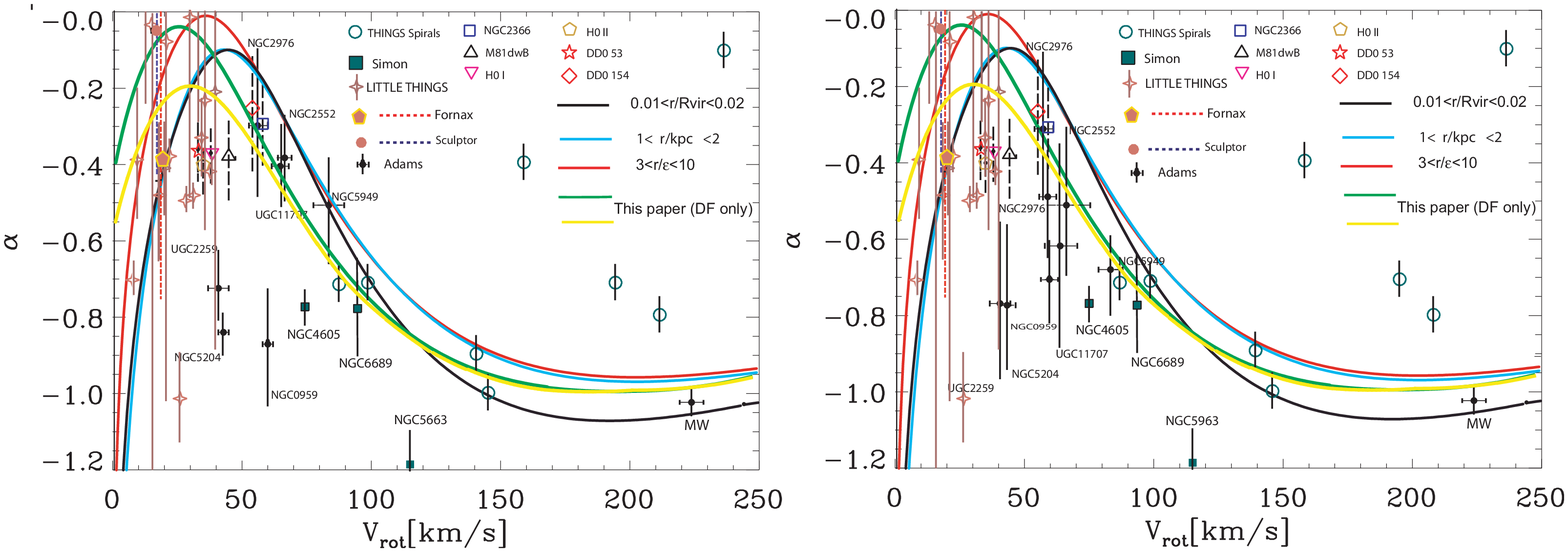}
 \caption{The $\alpha$-$V_{\rm rot}$ relation compared to observations. Data come from \protect\cite{Adams2014}, 
 \protect\cite{Simon2005} (excluding the galaxies studied by \protect\cite{Adams2014}), the THINGS spirals, the 
 THINGS dwarves, the LITTLE THINGS, Fornax, and Sculptor \protect\citep{Walker2011} and the MW 
 \protect\citep{Ascasibar2006}. The red, black, and blue lines describe the model of \protect\cite{DiCintio2014}, 
 while the green and yellow lines the model of the present paper. The only difference between the left and right 
 panel is due to the galaxy sample in \protect\cite{Adams2014}: gas traced, in the left panel; stellar traced, in 
 the right panel. The green (yellow) curves are our equivalent for the red (cyan) curve of 
 \protect\cite{DiCintio2014}.}
 \label{fig:alphavrot1}
\end{figure*}

In the last years, it has been shown that the SNFF model has some drawbacks. \cite{Penarrubia2012} calculated the 
energy that SN must inject into the haloes in order to remove the cusp. According to \cite{Penarrubia2012}, in 
order to transform a cuspy profile into a cored one in MW dSphs, an energy in the range $10^{53}-10^{55}$ erg is 
required. The average energy released in a SNII explosion is of the order of $10^{51}$ erg 
\citep{Utrobin2011}. So the explosions of hundreds to several thousands of SNs could in principle produce 
this huge amount of energy. However, the low star formation efficiency in dSphs, suggested by their luminous 
satellites abundance, implies that the real contribution to the energy could be lower than that needed to flatten a 
profile. Moreover, while the solution to the cusp/core problem with the SNFF model needs a large number of SNs, and 
so a large star formation efficiency (SFE), the solution of another small scale problem of the $\Lambda$CDM model, 
namely the TBTF problem, places an opposite demand on the SFE. 
In order to eliminate such a tension one or more of the following issues should be true: a) a coupling of energy 
coming from SN II to DM of the order of 1. Such a value of the "energy coupling" ($\epsilon_{\rm SN}$) contradicts 
observations. In order to describe the metallicity-luminosity relation in dSphs, \cite{Revaz2012} obtained a 
value of $\epsilon_{\rm SN} \simeq 0.05$, while the value of $\epsilon_{\rm SN}$ used in \cite{Governato2010} is 
0.40. Even this large value is smaller than what needed to eliminate the quoted tension.
b) Cusp removal at high redshifts ($z>6$ from the Sculptor and Fornax cored profiles). 
Namely, star formation should peak at redshifts unexpectedly high ($z>6$). 
This conclusion is at odds with the fact that star formation went on for 12-13 Gyr in Fornax \citep{deBoer2012b}, 
and for 6-7 Gyr in Sculptor \citep{deBoer2012a}. Even if the star formation peak was at that redshift, the 
tension moves to haloes embedding less stars, the formation of cores in dSphs with $M_{\ast}<10^7 M_{\odot}$ 
requires $\epsilon_{\rm SN} \simeq 1$. c) A top-heavy stellar initial mass function. d) Considerable satellite 
disruption (e.g., by tidal torques). This last issue is promising in the solution of the previously discussed problem 
as shown by \cite{Zolotov2012,DelPopolo2014b,Brook2014,DelPopolo2014d}.

\begin{figure}
 \centering
 \includegraphics[angle=-90,width=\hsize]{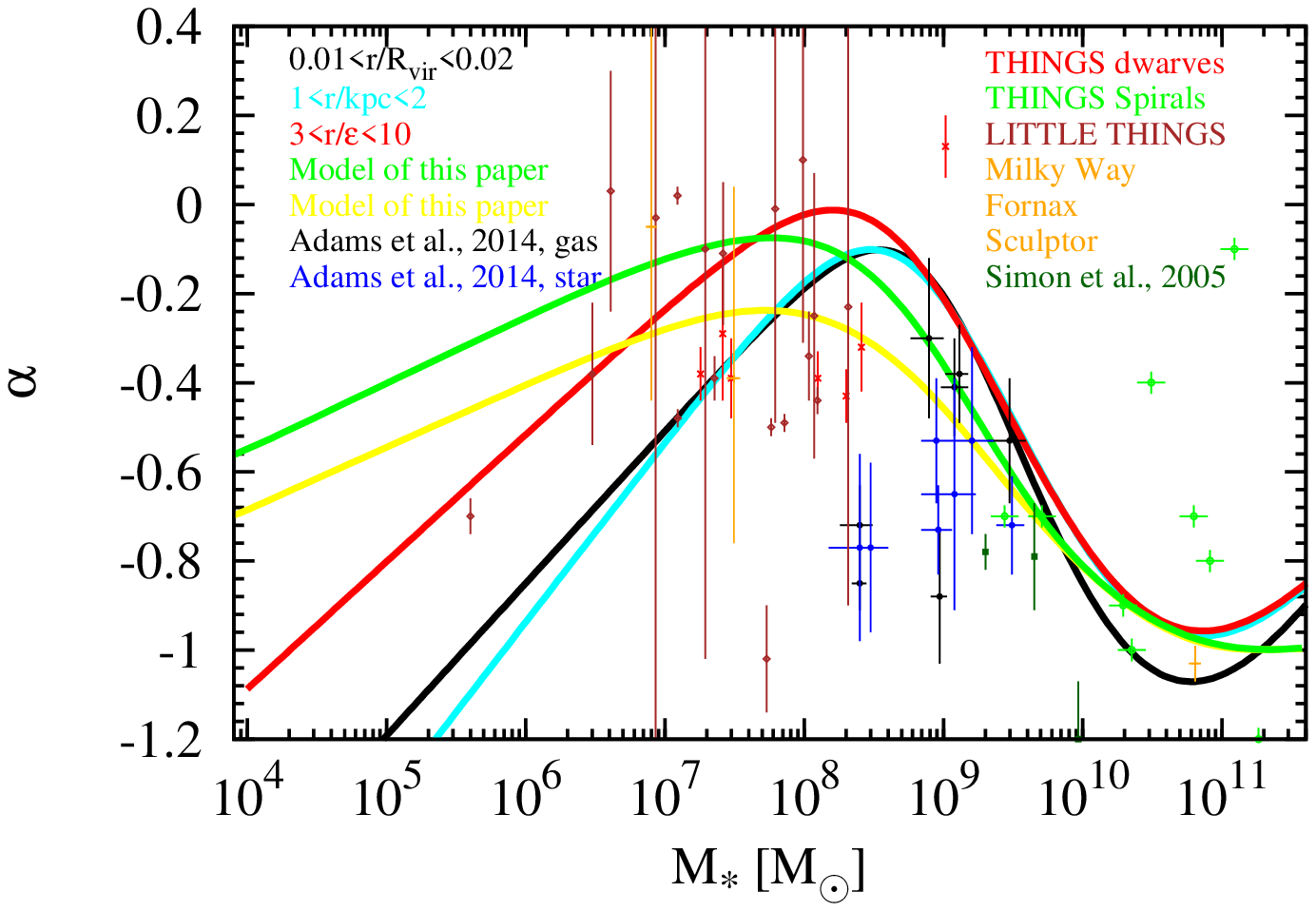}
 \caption[]{The $\alpha$ vs. $M_{\ast}$ relation. Symbols are like in figure~\ref{fig:alphavrot1}, {except for the 
 points representing the two samples from \protect{\cite{Adams2014}}: in black we show the gas-traced sample while in 
 blue the stellar-traced sample.}}
 \label{fig:alphaMstar}
\end{figure}

However, we should add that \cite{Governato2010} had to use a very high star formation threshold to obtain their 
results. \cite{Sawala2014} showed that the high threshold assumption is not necessary to obtain the results claimed by 
\cite{Governato2010}.

{One additional argument of discussion about simulations and the TBTF problem is given by \cite{Polisensky2014}. 
The authors could largely reconcile the discrepancy between simulations and observations by using the latest 
cosmological parameters provided by the WMAP and Planck team. They concluded therefore that the strong tension 
observed in previous works on the subject was mainly due to the incorrect assumption of the cosmological parameters. 
Another source of tension comes from the assumed mass of the Milky Way: assuming its most recent estimates, the need 
of baryonic physics necessary to decrease the density of the most massive satellites in the Milky Way becomes much less 
compelling. Despite this improvement, one more aspect lacks a satisfactory explanation, namely the problem of missing 
bright satellites just outside the Milky Way virial radius, as discussed in \cite{Bovill2011a}, \cite{Bovill2011b} and 
\cite{GarrisonKimmel2013}.}

On a similar line, \cite{GarrisonKimmel2013} discuss general problems of the SNFF model and in particular problems 
in solving the TBTF problem. Comparing the effects of blow-outs of different masses ($10^7 M_{\odot}$, 
$10^8 M_{\odot}$ and $10^9 M_{\odot}$) they found that a single blow-out of a fixed mass has more effect in changing 
the structure of a dwarf in comparison with several repeated blow-outs whose mass sums to the that of the single 
blow-out\footnote{However, repeated blow-outs remove mass preferentially from the centre.}, contrarily to the SNFF 
model which requires repeated blow-outs \citep[see][]{Governato2010,Governato2012,Pontzen2012}. From the 
point of view of the mass, for the high resolution simulations of \cite{GarrisonKimmel2013} to have subhaloes 
density in line with that observed in MW dSphs, a quantity of mass equal to $10^9 M_{\odot}$ should be ejected, 
which is marginally exceeding the baryon content of the dSphs. From the energy point of view, since the average 
energy emitted by SNs explosions is $10^{51}$ erg, to match the density of a $10^6 M_{\odot}$ dSph, the energy of 
40000 supernovae is needed with an efficiency of 100\%. For six of the nine classical dSphs, this quantity exceeded 
the number of Type II SN explosions expected. Moreover, they find that explosions can flatten the inner slope to 
$\alpha>-0.5$, never producing a real core with constant density, as predicted by the last versions of the SNFF model 
\citep[e.g.,][]{Pontzen2012}. \cite{Ferrero2012} and \cite{Papastergis2015} arrived to similar conclusions, namely 
the high improbability for the SNFF to solve the TBTF problem. Finally, as several authors noticed 
\citep{Choi2014,Laporte2015a,Laporte2015b,Marinacci2014}, nowadays hydrodynamical simulations have not the required 
resolution to follow the feedback processes which should transform the cusp into a core.

{Finally \cite{Oman2015} and \cite{Oman2016} with the use of hydrodynamical simulations investigate the `cusp 
versus core' problem and conclude that this is better characterized as an `inner mass deficit' problem rather than 
as a density slope mismatch. Investigating simulated dwarf galaxies and comparing their properties with a selected 
catalogue of observed galaxies, they find several discrepancies. The authors conclude that to solve these 
discrepancies, it is necessary either to change the dark matter physics, that the mass profiles of the galaxies giving 
rise to the `cusp versus core' problem are incorrect when inferred from kinematic data or that simulations fail to 
reproduce correctly observations, as pointed by other authors before.}

\begin{figure}
 \centering
 \includegraphics[angle=-90,width=\hsize]{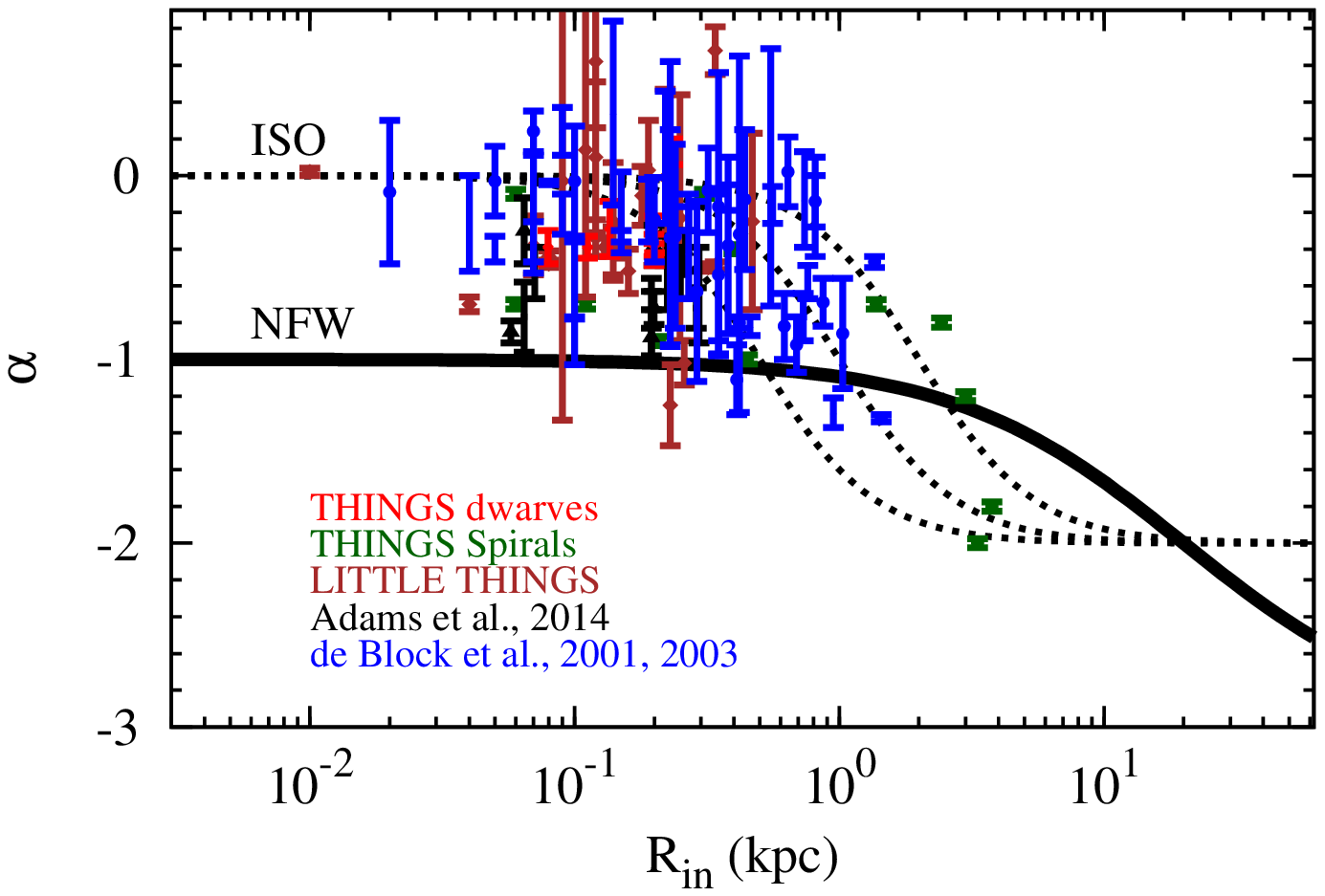}
 \caption{Inner slope vs. $R_{\rm in}$. The last quantity, the innermost radius point, is defined by means of a 
 least-square fit to the inner data point in the case of THINGS dwarves and LITTLE THINGS. In the 
 \protect\cite{Adams2014} case, $R_{\rm in}$ is obtained by adding the seeing to the fibre radius in quadrature. The 
 solid and dashed lines represents the theoretical prediction of NFW and ISO halos. The black symbols refer to the 
 \protect\cite{Adams2014} data, the 7 THINGS dwarves are indicated by the red symbols visible in the figure, and the 
 LITTLE THINGS are represented by the brown dots. {The THINGS SPIRALS are shown with the green points}. 
 The {\bf blue} symbols are the \protect\cite{deBlok2001} data 
 (open circles) and those of \protect\cite{deBlok2003} are shown with triangles.}
 \label{fig:alphaRin}
\end{figure}

\begin{figure}
 \centering
 \includegraphics[angle=-90,width=\hsize]{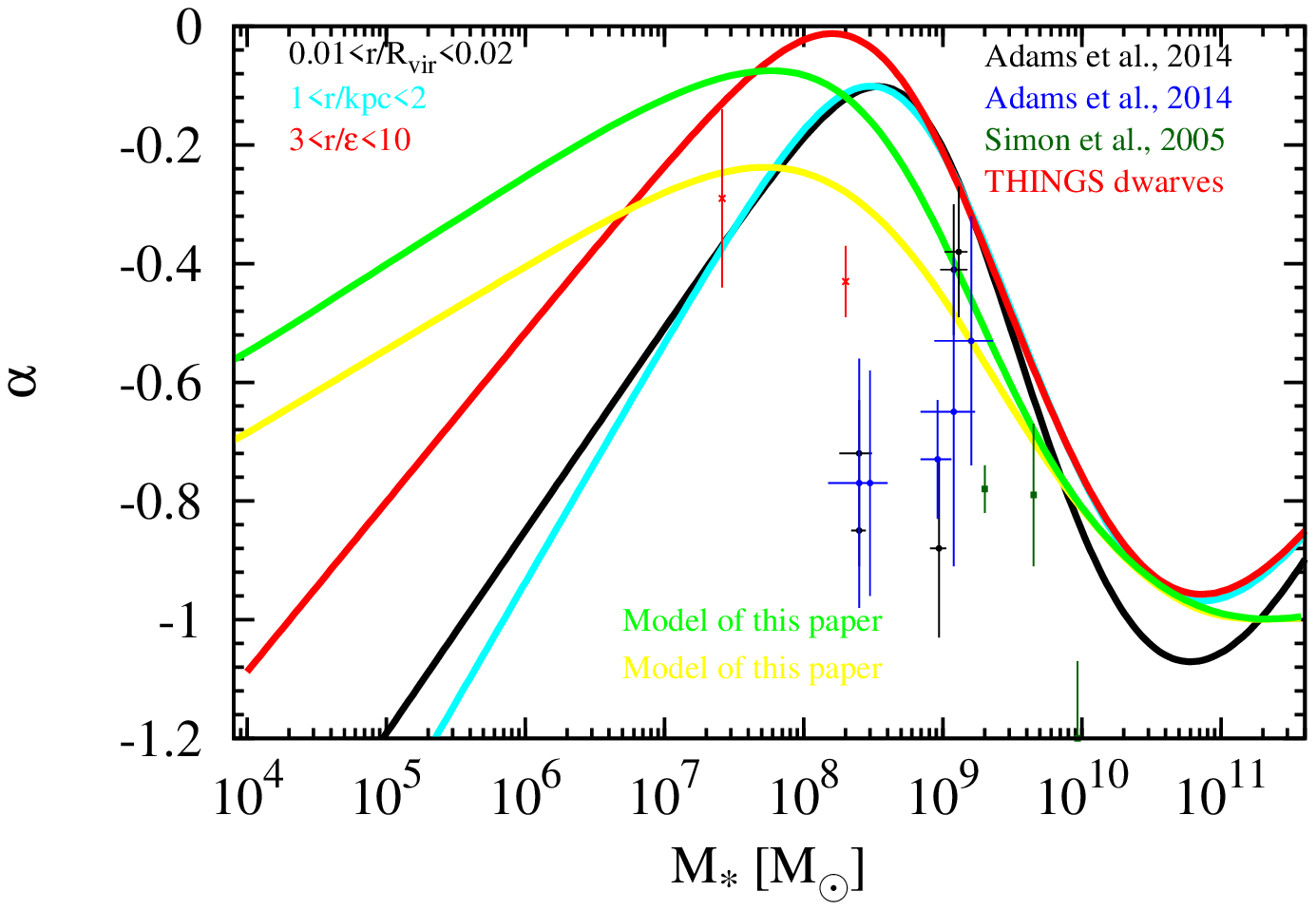}
 \caption[]{Like figure~\ref{fig:alphaMstar}, but now the data are only: \protect\cite{Adams2014} galaxy sample, 
 \protect\cite{Simon2005} galaxies (excluding those studied by \protect\cite{Adams2014}), and the best behaved 
 THINGS dwarves (Ho II and DDO 154).}
 \label{fig:alphaMstarOpt}
\end{figure}

Concerning the DFBC model, to our knowledge there are not studies on its limits. However, since the effectiveness of 
this process depends on the clumps properties and that of the halo, one could speculate about the a) the origin of 
the clumps, and b) the time-scales for the density flattening in comparison with the life of the clumps.

Concerning the first issue, some galaxies ("clump clusters" and "chain galaxies") at high redshift show clumpy 
structures \citep[e.g.][]{vandenBergh1996,Elmegreen2004,Elmegreen2009,Genzel2011}. 
At $t<2$ Gyr, when the gas is infalling towards the disc, radiative cooling induces a self-gravity instability which 
leads to clumps formation \citep[e.g.][]{Noguchi1998,Noguchi1999,Agertz2009,Bournaud2009,Ceverino2010}, and 
finally to the formation of disc galaxies.

Concerning the second issue, as shown by \cite{Nipoti2015}, the flattening process, $t_{\rm flat}$, happens on the 
dynamical friction scale $t_{\rm fric} \simeq 1.4 t_{\rm cross}/\ln{\Lambda}$, with $t_{\rm cross} \simeq 21$ Myr, 
namely the time-scale is very short.

At the same time, in order for the mechanism to work, the gas clumps should live for a time longer than that needed 
to redistribute DM in the halo centre. The evaluation of this life-time is not easy, since it is connected to a 
poorly known subject, namely the process of star formation and feedback 
\citep[e.g.][]{Murray2010,Genel2012,Hopkins2012,Bournaud2014}. The star formation time must be larger than 
$max[t_{\rm dyn}, t_{\rm cool}]$, where $t_{\rm dyn}=1/\sqrt{G \rho_{\rm gas}}$ is the dynamical time, and 
$t_{\rm cool}$ is the cooling rate. As estimated by \cite{Nipoti2015} (see their Sect. 3.3), this should be several 
tens of megayears. After star formation starts, if the clumps are destroyed by supernovae, we have to wait another 
$\simeq 10^7$ yr before the first star explodes.

For precision's sake, \cite{Genzel2011} observed outflows coming out from the stellar clumps parts of the clump 
clusters, and \cite{Genel2012} suggested that the clumps dissipate in a few tens of Myrs \citep{Murray2011}, to 100 
times this value in high-z clumps (compatible with the redshift of flattening). Moreover, \cite{Inoue2011} found that 
clumps are even more long-lived than what found in the previous papers.

Observation of the MW favours the long life thesis of clumps. Moreover, estimates of molecular clouds lifetime is 
$>10^8$ yrs \citep{Scoville2013}. According to the previous discussion, the clumps should orbit for several dynamical 
times before being dissipated.

\section{Results}\label{sect:results}
As previously discussed, there are at least two astrophysical solutions to the cusp/core problem: a) the SNFF 
mechanism, and b) the DFBC mechanism. In this section, we compare the data on the slopes of galaxies spanning a 
range of stellar masses $M_{\ast}=4\times 10^6 M_{\odot}-1.6 \times 10^{11} M_{\odot}$, with the two quoted models.

The results of the paper are presented in figures~\ref{fig:alphavrot}-\ref{fig:alphaMstarOpt}.

Figure~\ref{fig:alphavrot} shows the inner slope in terms of the rotation velocity of the total mass. The dashed lines 
reproduce the $\alpha$-$V_{\rm rot}$ relation calculated by \cite{DiCintio2014} for $\alpha$ measured in different 
ranges. The red line represents the range $3<r/\epsilon<10$, the blue line the range $1<r/{\rm kpc}<2$, and the black 
one, the range $0.01<r/R_{\rm vir}<0.02$, where $\epsilon$ is the softening length, and $R_{\rm vir}$ is the virial 
radius. In the case of the galaxies simulated by \cite{DiCintio2014}, the red line corresponds, for the lowest mass 
halo to a range $0.23~{\rm kpc}<r<0.78~{\rm kpc}$, and for the largest $0.94~{\rm kpc}<r<3.13~{\rm kpc}$. The black 
line $0.60~{\rm kpc}<r<1.20~{\rm kpc}$, and for the largest mass galaxy $1.30~{\rm kpc} <r<2.65~{\rm kpc}$.

The black line is fitted by \citep[see][]{DiCintio2014}:

\begin{equation}
\alpha_{\rm black}=0.132-\log{\left(\frac{\eta^{2.58}+1}{\eta^{1.99}}\right)}\;,
\end{equation}
with
\begin{equation}
\eta=0.84\left(\frac{M_{\ast}}{10^9 M_{\odot}}\right)^{-0.58}+
0.06\left(\frac{M_{\ast}}{10^9 M_{\odot}}\right)^{0.26}\;,
\end{equation}
while the blue and red lines are respectively fitted by
\begin{equation}
 \begin{split}
  \alpha_{\rm blue}=0.167619-\log\left[\left(10^{X+2.14248}\right)^{-0.699049}+\right.\\
  \left.\left(10^{X+2.14248}\right)^{1.56202}\right]\;,
 \end{split}
\end{equation}
and
\begin{equation}
 \begin{split}
  \alpha_{\rm red}=0.230967-\log\left[\left(10^{X+ 2.20929}\right)^{-0.493541}+\right.\\
  \left.\left(10^{X+2.20929}\right)^{1.49315}\right]\;,
 \end{split}
\end{equation}
with $X$ being
\begin{equation}
 X=\log{\left[\frac{0.0702}
 {\frac{10^{6.7120}M_{\ast}^{-0.57912}}{4.6570}+\frac{10^{-2.9657}M_{\ast}^{0.25589}}{0.50674}}\right]}\;.
\end{equation}

The equations for $\alpha_{\rm red}$ and $\alpha_{\rm blue}$ have been kindly provided by A. Di Cintio.

The previous equations give the $\alpha$-dependence on $M_{\ast}$. In order to get the dependence on $V_{\rm rot}$, 
we use (Eq.~\ref{eq:dutt}), the stellar mass Tully-Fisher (TF) relation (Eq.~4 of \cite{Dutton2010}), similarly to 
\cite{DiCintio2014}. Note that the three lines in figure~6 of \cite{DiCintio2014} and the three lines of this paper 
are different by a factor h inverse, coming from the Tully-Fisher relation of \cite{Dutton2010}, and discarded in 
\cite{DiCintio2014}.

In figure~\ref{fig:alphavrot1} we compared the theoretical relations obtained by \cite{DiCintio2014} and that 
obtained by means of our model with observational data. 
As before, the red, blue and black lines represent the slope $\alpha$ calculated at different distances from the 
galaxies centre by \cite{DiCintio2014}. The green line represents the same relation calculated with our model. 
They were calculated using the model in Appendix \ref{sect:app}, which is based on \cite{DelPopolo2009} 
using a different recipe of gas cooling, and taking into account star formation, reionization, and supernovae 
feedback as described in the Appendix \ref{sect:app} \citep[see also][]{DelPopolo2014a}. The model was used to study 
the evolution of proto-structures and formation of structures of different masses, with final halo mass, and stellar 
masses similar to that studied by \cite{DiCintio2014}. For each galaxy the stellar to halo ratio was calculated, and 
the slope calculated in the same radial bins of \cite{DiCintio2014} (e.g., see their figure~3). Converting the stellar 
mass into halo mass by means of the Moster relation \citep{Moster2013}, we can calculate the $\alpha$-$M_{\ast}$ 
relation. This can be converted into an $\alpha$-$V_{2.2}$ relation\footnote{$V_{2.2}$ is the rotation velocity for 
late-type galaxies, at 2.2 disc scale-lengths. $V_{2.2} \simeq V_{\rm opt}$ for the quoted late-type galaxies.}, 
through the Tully-Fisher relation \citep[Eq. 4]{Dutton2010}. 
Our green (yellow) line corresponds to the red (blue) one in \cite{DiCintio2014}, their figure~6. Note that the turn 
in the $\alpha$-$V_{\rm rot}$ relation at $\simeq 25$ km/s (in the case of our model) originates from the fact that 
in our model the inner slope of the density profile of a structure is inversely proportional to the angular momentum 
\citep[e.g.][]{DelPopolo2009}: the larger the last the flatter is the profile. When we move from normal spiral 
galaxies to the dSphs region the angular momentum of the structure strongly decreases producing the steepening of the 
inner slope.

These models are compared with a set of data composed by the \cite{Adams2014} galaxies, the \cite{Simon2005} 
galaxy sample \citep[excluding those restudied by][]{Adams2014}, the THINGS dwarves \citep{Oh2011a,Oh2011b}, the 
THINGS galaxies \citep{Chemin2011,Oh2015}, the LITTLE THINGS galaxies, provided by Oh, and Fornax, and Sculptor, 
whose inner slope was calculated by \cite{Walker2011} without adopting a DM halo model, and directly from stellar 
spectroscopic data. They measured the quantity $\Gamma\equiv\Delta\log{M}/\Delta\log{r}$, finding a 
value of $\Gamma=2.61_{-0.37}^{+0.43}$ for Fornax, and $\Gamma=2.95_{-0.39}^{+0.51}$ for Sculptor. 
The relation among $\Gamma$ and $\alpha$ is given by $\alpha_{\rm DM}< 3-\Gamma$ \citep{Walker2011}. We added also a 
point coming from MW, obtained from the best fit to the 511 keV emission \citep{Ascasibar2006}. The figure shows 
that for $V_{\rm rot}>100$ km/s, only 4 points over 9 are compatible with the theoretical models. For smaller values 
of the velocity, the \cite{Adams2014} galaxy sample shows a larger slope in comparison with those of other samples 
(as observed by the same authors). The majority of the THINGS dwarf galaxies have slope larger than the predictions. 
The LITTLE THINGS data can be used to constrain the models only at small velocities ($<40$ km/s). They agree with 
the theoretical data, similarly to Sculptor and Fornax, but they have large errors. The plot shows that the main 
differences among the SNFF and the DFBC model are evident in the velocity range $50-100$ km/s. The DFBC predicts 
steeper slopes, with maximum differences around $\Delta\alpha\simeq 0.2$. The other difference is evident 
at small velocities. While the SNFF predicts slopes that steepen to cuspy, since at small velocities corresponding to 
$M_{\ast}/M_{\rm halo} \leq 0.01$\% the energy from supernovae feedback is not enough to flatten the profile, the 
DFBC predicts flatter profiles. The reason is connected to the different mechanism the DFBC is based on. As already 
reported, in dSphs star formation efficiency is low. This means that if we have a fixed quantity of gas, the clumps 
that it forms can act directly on DM, while in the SNFF mechanism the gas must be converted into large mass stars 
(and as we told the efficiency is low) to explode into SNs that will inject energy in the DM.

The ability of the DFBC mechanism to form profiles flatter than the SNFF mechanism is important since it implies 
that dSphs of low mass are not necessarily cuspy, as predicted by the SNFF model. This means that the observation of 
cored profiles at small stellar masses (e.g., $<10^6 M_{\odot}$) would not imply that the $\Lambda$CDM model has 
scarce possibilities to be correct, as it is predicted by the SNFF model \citep[e.g.,][]{Madau2014}.

In the left panel of figure~\ref{fig:alphavrot1}, we plot the same quantities, except that the Adams' data are 
obtained from the stellar traced observations.

Figure~\ref{fig:alphaMstar} shows $\alpha$ in terms of the stellar mass $M_{\ast}$. The two figures give similar 
information as the two panels in figure~\ref{fig:alphavrot1}. However, in this case we did not use Eq.~\ref{eq:dutt} 
to transform $M_{\ast}$ into $V_{\rm rot}$ with the result that we have smaller uncertainties. Moreover, the plots 
show more clearly the behaviour of the models especially at small stellar masses. The plots show that the DFBC model 
predicts steeper profiles in the range $10^8 M_{\odot}<M_{\ast}<10^{10} M_{\odot}$ with respect to the SNFF model. 
At small stellar masses, $M_{\ast}\simeq 10^4 M_{\odot}$ the profile produced by the DFBC is not cuspy, at least it 
does not become steep as a NFW or steeper as happens to the SNFF model. Apart the visual difference among the two 
models, we have analysed which one describes better the data. We applied a $\chi^2$ to the data and model (see the 
following subsection).

Since data having not sufficient spatial resolution can give rise to larger values of $\alpha$, in cored systems, it 
is useful to study the behaviour of the logarithmic slope in terms of the data spatial resolution. In 
figure~\ref{fig:alphaRin}, we plot the logarithmic inner slope as a function of the RC's resolution. 
As \cite{Oh2011a}, we plot data coming from THINGS dwarves, \cite{deBlok2001} (open circles), \cite{deBlok2003} 
(triangles). We add data from \cite{Adams2014} and the LITTLE THINGS objects. As the figure shows, at low resolution 
$R_{\rm in} \simeq 1$ the NFW and ISO slopes are almost equal \citep{deBlok2001}, while at high 
resolution, $R_{\rm in}<1$, it is possible to distinguish the NFW and ISO slopes. The solid and dashed curves 
represent the theoretical NFW and ISO $\alpha$-$R_{\rm in}$ relations. While the THINGS dwarves and the LITTLE THINGS 
deviate significantly from the NFW predictions, the \cite{Adams2014} data are somehow intermediate between the two 
models, namely intermediate between cuspy and cored halos.

In the following, in order to constrain the two astrophysical models able to transform cuspy profiles in cored ones, 
we will first use a) all the galaxies previously discussed, and then in order to get the best available distribution 
of slopes, and constraints on the two quoted mechanisms, we will b) include only the highest-quality results 
available in literature. We will then include the sets by \cite{Adams2014} and \cite{Simon2005} not re-studied in 
\cite{Adams2014} and some of the THINGS galaxies. 

Concerning the case b, we have to choose a sample. From the previous discussions, we know that the data sets are 
noteworthy different. The average values of $\alpha$ is different for different data sets. The THINGS dwarf galaxies 
have $<\alpha>=-0.29 \pm 0.07$ \citep{Oh2011b}, which is significantly shallower than the \cite{Adams2014} data set 
having $<\alpha>=-0.67 \pm 0.10$ (stellar kinematics), and $<\alpha>=-0.58 \pm 0.024$ (gas), or those of 
\cite{Simon2005} with $<\alpha>=-0.73 \pm 0.44$.

The question is what originates those differences. 
The difference among the THINGS dwarves and the \cite{Adams2014} result could be due to the fact that the THINGS 
dwarves have smaller stellar masses. However, this point seems not to be so important, because the stellar mass of 
the THINGS dwarves is enough to give rise to a core in the quoted galaxies, according to the models of SN feedback. 
\cite{Oh2011b} measured the inner slopes by using a power-law fit to the innermost $\simeq 5$ points of the 
rotation curve ($r \simeq 1$ kpc). These points are obviously the most exposed to systematic uncertainties. 
The slopes measured in \cite{Adams2014} were calculated by fitting the entire density profile. Moreover, a large part 
of the THINGS dwarves have evident peculiarities that could bias the slopes measurement
\footnote{Holmberg I, Holmberg II, NGC 2366, DD0154, and DDO 53 have not trivial kinematic asymmetries, IC 2574, 
NGC 2366 have not negligible non-circular motions, and M81dwB has a rotation curve which declines.}.

The THINGS galaxies have some drawbacks. \cite{deBlok2008} didn't derive the slopes of the THINGS galaxies as 
(for the spirals) these would be dominated by the stars, and corrections for them would be too uncertain. 
The only ones where they thought the slope would say something about the DM were the ones published in 
\cite{Oh2011a,Oh2011b} (which are dwarves).

Other authors \citep{Chemin2011}, tried to calculate the slopes of the THINGS galaxies. 
They re-analysed 17 galaxies, undisturbed and rotationally dominated, in the \cite{Walter2008} sample that coincides 
fundamentally with that of \cite{deBlok2008} and found that the mass distributions differ from those of 
\cite{deBlok2008}.

Also \cite{Oh2015} calculated the THINGS slopes, and the results are in many cases in conflict with those of 
\cite{Chemin2011}. In the present paper, we use \cite{Oh2015} data for the THINGS galaxies, {\it
for compatibility reasons}, since we also used Oh's data for THINGS dwarves and LITTLE THINGS. The reason behind 
it is that it is important to use galaxies whose slope is determined in a similar way and approximately at the same 
distance from the centre. Hence our choice for using data from \cite{Oh2015} and \cite{Adams2014} and neglecting 
those from \cite{Chemin2011}.

This justifies the need to choose accurately the best sample from the data we have. 

So, we will include the best behaved THINGS dwarves (Ho II and DDO 154), the Adams' galaxies (NGC 959; UGC 2259; 
NGC 2552; NGC 5204; UGC 11707) and those of \cite{Simon2005} not re-studied in Adams. This set has a 
$\alpha=-0.673 \pm 0.34$ almost independent of whether we use gaseous or stellar kinematics, and in strict 
agreement with the slopes obtained by \cite{Newman2013a} (see also Del Popolo 2014).

In figure~\ref{fig:alphaMstarOpt}, we compare the $\alpha$-$M_{\ast}$ relations with the previous sample.  

As reported in the introduction, the cusp/core problem has been also noticed at clusters of galaxies scales. 
It is remarkable that the mechanism explaining the quoted observations, and the correlations found by 
\cite{Newman2013a,Newman2013b}, described in \cite{DelPopolo2012c} and \cite{DelPopolo2014}, is the same explaining 
the shallow density profiles over 6 order of magnitudes in the halo mass (dwarfs, clusters). 
In order to explain the galaxy clusters density profiles, \cite{Martizzi2012} had to invoke the feedback from 
active galactic nuclei (AGN) on the distribution of gas, finding a cores of similar size in the DM component and 
the stellar central component.


\begin{table*}
 \centering
 \small
 \caption[reduced chi square]{$\chi^2_{\rm red}$ values for different samples and models.\label{tab:tabchi2r}}
 \begin{tabular}{lccccc}
 \tableline
 Sample              & $0.01<r/R_{\rm vir}<0.02$ & $1<r/{\rm kpc}<2$ & $3<r/\epsilon<10$ & Model of this paper & 
                       Model of this paper \\
                     & & & & for $1<r/{\rm kpc}<2$ & for $3<r/\epsilon<10$ \\
 \tableline
 All (gas)           & 139.95 & 144.16 & 153.30 & 128.12 & 150.85 \\
 All (stars)         & 136.71 & 140.88 & 149.15 & 125.81 & 147.61 \\
 Optimal (gas)       & 39.67  & 41.47  & 49.13  & 20.78  & 30.66 \\
 Optimal (stars)     & 21.01  & 22.46  & 25.39  & 8.18   & 12.79 \\
 \tableline
 $V_{\rm rot}<50$~km/s (gas)   & 53.98 & 59.85 & 77.28 & 41.71 & 82.59 \\
 $V_{\rm rot}<50$~km/s (stars) & 47.20 & 52.96 & 68.74 & 37.28 & 76.30 \\
 \tableline
 $V_{\rm rot}<30$~km/s         & 70.00 & 78.99 & 93.76 & 57.72 & 114.36 \\
 \tableline
 \end{tabular}
\end{table*}

\subsection{Determination of the best model}
As shown in the previous figures, the different models do not seem, as one may expect, to reproduce the data 
perfectly. 
This is understandable since on one side the theoretical models are very approximate and suffer of many uncertainties 
due to the poor knowledge of the baryon physics involved, while on the other side, observational mass and inner slope 
determinations are affected by systematics and difficulties.

Therefore it is necessary to adopt a statistical approach to evaluate which model describes better the observations. 
To do so we use the $\chi^2$ test defined as
\begin{equation}
 \chi^2=\sum_{k=1}^{N}\frac{(\alpha_{\rm th}-\alpha_{{\rm obs},k})^2}{\sigma^2_{k}}\;,
\end{equation}
where $\alpha_{\rm th}$ and $\alpha_{\rm obs}$ are the theoretical and observed density slope respectively and 
$\sigma^2_{k}$ the error on each measurements. This test strictly requires that the error distribution is Gaussian. 
If this is not the case (being the error bars asymmetric), we will use the geometric mean of the two errors. Note 
also that we do not have any free parameter in the models, being the models dependent only on the mass of the stellar 
component $M_{\ast}$ and the numerical constants determined by the underlying physical processes considered.

We will build several likelihood functions: we first consider the whole sample of data points and then the restricted 
subsample described in Section~\ref{sect:results}.

In a second step we will consider again all the data points, but we will apply a velocity cut-off. In particular, we 
will select all the points with maximum velocity $V_{\rm rot}$ smaller than 50 and 30 km/s. We do not go below this 
threshold since there would be not enough points. With a threshold of 50 (30) km/s we can use 27 (16) data points out 
of the 50 used in this work. The optimal sample is made of 10 objects.

{We show our results about the reduced chi-square ($\chi^{2}_{\rm red}$) in table~\ref{tab:tabchi2r} using the 
numerical and theoretical values of the inner slope $\alpha$ as in figure~\ref{fig:alphavrot1}. 
In this way we do not need any conversion from the velocity $V_{\rm rot}$ to the stellar mass $M_{\ast}$. 
Note that the values of $\chi_{\rm red}^2$ are very high, of the order of hundred per degree of freedom, due to the 
fact that many points lie outside the theoretical curves for many $\sigma$s and the $\chi^2$ statistics is dominated 
by points with very small error bars. We remind the reader that the reduced $\chi^2$ is defined as 
$\chi^2_{\rm red}=\chi^2/\nu$ where $\nu$ is the number of degrees of freedom of the model.}

As it appears evident, all the models perform very similarly, with marginal differences not statistically significant 
when we use all the data points, either with gas or stellar tracers. The situation is different when we restrict our 
analysis to the optimal sample. In this case the $\chi^2_{\rm red}$ for the sample containing objects whose inner 
slope was evaluated using stellar tracers is two times smaller than the one for the gas tracer sample. This shows the 
higher quality of the observations. We observe this improvement only for the optimal sample since when we apply a 
cut-off of 50~km/s in $V_{\rm rot}$, the two reduced $\chi^2$ are once again very similar. It is also instructive to 
compare the two different models studied in this work. Comparing the models' prediction for the same range of $r$, we 
observe that when using the whole sample, the two different models behave roughly the same, with no particular 
preference for one or the other. Same conclusion when we consider the cut $V_{\rm rot}<30$~km/s, with the model of 
this paper performing a bit better (worse) for the range $1<r/{\rm kpc}<2$ ($3<r/\epsilon<10$). When we restrict the 
analysis to the optimal sample, the model of this paper has a reduced $\chi^2_{\rm red}$ a factor of two smaller than 
the model based on the SN feedback, showing therefore a better fit to the data points.

{A usual quantity that can be easily evaluated the relative probability of the models with respect to each other. 
This is defined as the ratio of the likelihood of the models, and in terms of the $\chi^2$ statistics is expressed as 
$P(a,b)=\exp{[-(\chi_a^2-\chi_b^2)/2]}$ where $a$ and $b$ represent the two models, the SNFF and the DFBC, 
respectively. 
In other words we are asking which model is more likely to fit the data. We show our findings in table~\ref{tab:prob}. 
As expected the values are either very small or very big, due to the fact that the $\chi^2$ is significantly different 
from one. This is one more confirmation that the models fit the current data very poorly, also for the optimal sample, 
despite a significant improvement in terms of probability with respect to the full sample.}

\begin{table*}
 \centering
 \small
 \caption[probability]{Relative probability between the models for different samples.\label{tab:prob}}
 \begin{tabular}{lccc}
  \tableline
  Sample              & $1<r/{\rm kpc}<2$      & $3<r/\epsilon<10$ \\
  \hline
  All (gas)           & $1.83\times 10^{-171}$ & $7.97\times 10^{-27}$ \\
  All (stars)         & $5.37\times 10^{-161}$ & $4.15\times 10^{-17}$ \\
  Optimal (gas)       & $3.63\times 10^{-41}$  & $8.19\times 10^{-37}$ \\
  Optimal (stars)     & $1.26\times 10^{-28}$  & $2.33\times 10^{-25}$ \\
  \tableline
  $V_{\rm rot}<50$~km/s (gas)   & $3.67\times 10^{-103}$ & $9.39\times 10^{29}$ \\
  $V_{\rm rot}<50$~km/s (stars) & $2.89\times 10^{-89}$ & $4.89\times 10^{42}$ \\
  \tableline
  $V_{\rm rot}<30$~km/s         & $5.22\times 10^{-70}$ & $1.27\times 10^{67}$ \\
  \tableline
  \end{tabular}
\end{table*}

We can therefore conclude that while in general the two different models studied in this work perform similarly, when 
we restrict to a subsample of data, the model of this paper based on dynamical friction performs better than the model 
based on SN feedback {even if not at an appreciable statistically significant level.}

\section{Conclusions}\label{sect:conclusions}

The aim of this work is to compare the predictions of two models thought to be able to solve the cusp/core problem 
\citep{Moore1994,Flores1994}. The models taken into account are the ones based on the SN feedback (see 
Sect.~\ref{sect:sff}) and on the baryon clumps (see Sect.~\ref{sect:gcm}).

The first one assumes as main driver of the flattening the internal density profile, the action of the SN explosions 
on the surrounding medium causing the expulsion of gas, while the second one is based on the idea that energy and 
angular momentum transfer from baryon clumps to the dark matter component can flatten the profile and transform a 
cusp into a core.

We compare the models with a sample of observational data built from different sets (see Sect.~\ref{sect:data} for a 
detailed description). Here we limit ourselves to a short description.

One of the catalogues adopted is based on the sample of seven nearby galaxies by \cite{Adams2014} using both gas and 
stars as tracers, together with three additional objects studied by \cite{Simon2005}. A second data set is based on 
seven THINGS dwarves studied by \cite{Oh2011a,Oh2011b} and a third sample is based on the twenty-one objects of the 
LITTLE THINGS set, kindly provided by Se-Heon Oh \citep{Oh2015}. A fourth set, also provided by Se-Heon Oh 
\citep{Oh2015} includes the THINGS spirals and finally, the last set is built with data for the Milky Way, Fornax and 
Sculptor galaxies.

Analysing figures~\ref{fig:alphaMstar} and~\ref{fig:alphaMstarOpt} (the first with the full set of data points, the 
second with a restricted subsample made of the \cite{Adams2014} and \cite{Simon2005} galaxies, and the two THINGS 
dwarves Ho II and DDO 154) and comparing the models with a reduced chi-square $\chi^2_{\rm red}$ analysis, we see 
that the whole data set does not favour any of the two models in particular, while a restricted analysis to the 
mentioned higher quality subsample shows a clear preference for the model based on baryon clumps.

We also showed that dSphs of small mass ($M_{\ast}<10^6 M_{\odot}$) can have cored profiles. For 
$M_{\ast}<10^5 M_{\odot}$ the slope is $\simeq -0.6$. This is the main difference between the SNFF model and the 
DFBC of this paper. An important consequence is that finding a dSphs having $M_{\ast}<10^6 M_{\odot}$ with an inner 
profile not cuspy, as predicted by the SNFF model, is not death hit for the $\Lambda$CDM model. 

The last point clearly shows that the determination of the inner structure of dwarf galaxies is of fundamental 
importance to determine the nature of the DM. However, as already reported the inner structure of DM haloes of dSphs 
is still debated, since it is difficult to distinguish cuspy and cored profiles \citep[e.g.,][]{Strigari2014}. 
Can this problem been solved in the near future? Some authors hinted to this possibility. 
Future observations from the Subaru Hyper-Supreme-Camera \citep{Takada2010} or GAIA \citep{deBruijne2012} have 
been indicated as a possible way out from the puzzle. In reality, even with those observations the problem will 
not be solved except for some larger dwarf galaxies \citep[e.g. Sagittarius, see][]{Richardson2014}. In fact, as 
previously discussed, one method often used to study the density profiles is based on the Jeans equations. The method 
has a drawback, a degeneracy between the density profile and the anisotropy parameter $\beta$. The direct 
determination of this parameter is not possible having just data on the 2D projection of stars radius and from the 
line of sight component of stars velocity. Several improvements to the previous case have been proposed 
\citep[see][]{Battaglia2013,Richardson2014}. Better results could be obtained by means of the 2+1 data sets (meaning 
that we know two of the three position coordinates and one of the three velocity coordinates) 
\citep[see][]{Battaglia2013}. 
Information on the proper motions allows the determination of the density slope at half-light radius 
\citep{Strigari2007}. However, the proper motions of dwarf galaxy stars is challenging to determine even with GAIA 
\citep{Richardson2014}, with a maximum astrometric accuracy of 7 $\mu$as at magnitude $V=10$, while it could have 
been possible with an error of $\pm 0.2$, determining the proper motions of just 200 stars, with the 
SIM mission\footnote{The Space Interferometry Mission, or SIM was a planned space telescope developed by USA. 
SIM was postponed several times and finally cancelled in 2010.}, which should have had a higher astrometric accuracy 
than GAIA's \citep{Strigari2007}.

\section*{Acknowledgements}
The authors thank Joshua Simon and Arianna Di Cintio for useful discussions and Se-Heon Oh for kindly providing his 
data points.

\appendix

\section{Model}\label{sect:app}

The model used in the present paper was introduced in \cite{DelPopolo2009,DelPopolo2014a}, and then applied in 
several other papers to study the universality of the density profiles \citep{DelPopolo2010,DelPopolo2011}, the 
density profiles in galaxies \citep{DelPopolo2012a,DelPopolo2014} and clusters \citep{DelPopolo2012b,DelPopolo2014}, 
and the inner surface-density of galaxies \citep*{DelPopolo2013d}.

The model is a semi-analytical model (SAM) that includes an improved secondary infall model (SIM) 
\citep[e.g.,][]{Gunn1972,Hoffman1985,DelPopolo1997,Ascasibar2004,Williams2004,Hiotelis2006,Hiotelis2013,Cardone2011a,
DelPopolo2013a,DelPopolo2013b}. Differently from previous SIMs, the model considers the effects of non-radial collapse 
originated by random angular momentum
\footnote{The random angular momentum arises from the system random velocities \citep{Ryden1987}.}, 
adiabatic contraction of DM that baryons produce 
\citep{Blumenthal1986,Gnedin2004,Gustafsson2006}, dynamical friction effects \citep[e.g.,][]{ElZant2001,ElZant2004}, 
random and ordered angular momentum. The model takes into account reionization, cooling, star formation, and the 
supernova feedback (see the following).

The model follows the evolution of a perturbation starting from the linear phase, expanding with the Hubble flow till 
the phase of maximum expansion (turn-around). In the following phases of collapse and "shell-crossing", it is 
assumed that the central potential varies adiabatically \citep{Gunn1977,Fillmore1984a}. 
The final profile is given by
\begin{equation}\label{eq:dturnnn}
 \rho(x)=\frac{\rho_{\rm ta}(x_{\rm m})}{f(x_{\rm i})^3}
 \left[1+\frac{d\ln{f(x_{\rm i})}}{d\ln{g(x_{\rm i})}}\right]^{-1}\;,
\end{equation}
where we indicated the initial radius with $x_{\rm i}$, the collapse factor with 
 $f(x_{\rm i})=x/x_{\rm m}(x_{\rm i})$, and the turn-around radius with $x_{\rm m}(x_{\rm i})$, given by
\begin{equation}
 x_{\rm m}=g(x_{\rm i})=x_{\rm i}\frac{1+\overline{\delta}_{\rm i}}
 {\overline{\delta}_{\rm i}-(\Omega_{\rm i}^{-1}-1)}\;.
\end{equation}

In the previous equation, $\Omega_{\rm i}$ is the density parameter, and $\overline{\delta}_{\rm i}$ the average 
overdensity in a given shell. 
Our model contains DM and baryons. Initially, baryons are in the gas phase. Baryon fraction is set equal to the 
"universal baryon fraction" $f_{\rm b}=0.17\pm 0.01$ \citep{Komatsu2009} \citep[0.167 in][]{Komatsu2011}. 
The baryonic fraction is obtained from the star formation processes described in the following.

In the model, the "ordered angular momentum" $h$ (coming from tidal torques of large scale structures on those on 
the smaller scales) is obtained through the tidal torque theory (TTT) 
\citep{Hoyle1953,Peebles1969,White1984,Ryden1988,Eisenstein1995}. 
The "random angular momentum" $j$ is expressed in terms of the eccentricity 
$e=\left(\frac{r_{\rm min}}{r_{\rm max}}\right)$ \citep{AvilaReese1998}, where $r_{\rm max}$ is the apocentric 
radius, and $r_{\rm min}$ the pericentric radius. 
A correction on the eccentricity is made, according to simulations of \cite{Ascasibar2004}, which consider the 
effects of the dynamical state of the system on eccentricity
\begin{equation}
e(r_{\rm max})\simeq 0.8\left(\frac{r_{\rm max}}{r_{\rm ta}}\right)^{0.1}\;,
\end{equation}
for $r_{\rm max}<0.1 r_{\rm ta}$.

The steepening of the profile produced by the adiabatic compression was obtained following \cite{Gnedin2004}, 
and calculated using iterative techniques \cite{spedicato},
while the effects of dynamical friction were obtained adding the dynamical friction force, calculated as described in Appendix A of \cite{DelPopolo2009}, to the equation of motions \citep[see][Eq. A14]{DelPopolo2009}.

Gas cooling, star formation, reionization and supernovae feedback were included as done by \cite{DeLucia2008} and 
\cite{Li2010} (Sect.~2.2.2 and~2.2.3).

Reionization, treated as in \cite{Li2010}, reduces the baryon content, and the baryon fraction changes as
\begin{equation}
 f_{\rm b, halo}(z,M_{\rm vir})=\frac{f_{\rm b}}{[1+0.26 M_{\rm F}(z)/M_{\rm vir}]^3}\;,
\end{equation}
where $M_{\rm F}$, is the "filtering mass" \citep*[see][]{Kravtsov2004}, and as usual the virial mass is indicated as 
$M_{\rm vir}$. The reionization redshift is in the range 11.5-15.

Gas cooling is treated as a classical cooling flow \citep[e.g.,][]{White1991} \citep[see Sect.~2.2.2 of][]{Li2010}. 
Similar results are obtained using the \cite{Ryden1988} treatment.

The details of star formation are given in \cite{DeLucia2008}. The treatment of \cite{Croton2006} is used for the 
supernovae feedback. In \cite{DiCintio2014} it was used the blast wave SN feedback \citep{Stinson2006}. 
For our purposes, the choice of the formalism, even if similar, is not so fundamental. A fundamental difference 
among our model and the SNFF model \citep[e.g.,][]{DiCintio2014} is that in our case the flattening process happens 
before star formation, and the source of energy is gravitational. Stellar feedback acts when the core is already 
formed, and disrupts the gas clouds. 
In the SNFF model the flattening process happens after star formation and the source of energy is stellar feedback.

Concerning these last steps. Gas forms a disc, and the star formation rate is
\begin{equation}
\psi=0.03 M_{\rm sf}/t_{\rm dyn}\;,
\end{equation}
being $t_{\rm dyn}$ the disc dynamical time, and $M_{\rm sf}$ the gas mass above a given density threshold, 
$n>9.3/cm^3$ as in \cite{DiCintio2014}. The initial mass function (IMF) is a Chabrier one \citep{Chabrier2003}. The 
amount of stars forming 
is given by
\begin{equation}
 \Delta M_{\ast}=\psi\Delta t\;,
\end{equation}
where $\Delta t$ indicates the time-step.

The quantity of energy injected by SN in the ISM is
\begin{equation}
 \Delta E_{\rm SN}=0.5\epsilon_{\rm halo}\Delta M_{\ast} V^2_{\rm SN}\;,
\end{equation}
where $V^2_{\rm SN}=\eta_{\rm SN}E_{\rm SN}$ is the energy injected per supernovae and per unit solar mass. 
The efficiency reheating disc gas efficiency produced by energy is fixed at $\epsilon_{\rm halo}=0.35$ 
\citep{Li2010}. $\eta_{\rm SN}=8\times 10^{-3}/M_{\odot}$ gives the supernovae number per solar mass obtained 
assuming a Chabrier IMF \citep{Chabrier2003}, and the typical energy released in a SN explosion is 
$E_{\rm SN}=10^{51}$ erg.

Energy injection in the gas reheats it proportionally to the number of star formed
\begin{equation}
 \Delta M_{\rm reheat} = 3.5 \Delta M_{\ast}\;.
\end{equation}

The change in thermal energy produced by the reheated gas is given by
\begin{equation}
 \Delta E_{\rm hot}= 0.5\Delta M_{\rm reheat} V^2_{\rm vir}\;,
\end{equation}
This hot gas will be ejected by the halo if $\Delta E_{\rm SN}>\Delta E_{\rm hot}$, and the quantity is
\begin{equation}
 \Delta M_{\rm eject}=\frac{\Delta E_{\rm SN}-\Delta E_{\rm hot}}{0.5 V^2_{\rm vir}}\;.
\end{equation}

The halo can accrete the ejected material that becomes part of the hot component related to the central galaxy 
\citep{DeLucia2004,Croton2006}.

The results of the previous model are in agreement with previous studies on the cusp flattening produced by heating 
of DM by collapsing clumps of baryons \citep{ElZant2001,ElZant2004,RomanoDiaz2008,Cole2011,Inoue2011,Nipoti2015}, and 
as previously reported predicted the correct shape of galaxy, and clusters density profiles together with a series of 
correlations found in observations like those of \cite{Newman2013a,Newman2013b}.

\end{document}